\documentclass[aps, prb, preprint]{revtex4-1}
\pdfoutput=1
\usepackage{mathtools}
\usepackage{graphicx}
\usepackage{color}
\usepackage{bm}
\usepackage{dcolumn}
\usepackage{comment}
\usepackage{subcaption}
\usepackage[strings]{underscore}
\usepackage{float}
\usepackage{textcomp}
\usepackage{amsmath}
\usepackage[font=sf]{caption}
\usepackage[normalem]{ulem}

\AtBeginDocument{}
\renewcommand{\figurename}{\textbf{Figure}}
\begin{document}
\def\be{\begin{equation}}
\def\ee{\end{equation}}
\def\beq{\begin{eqnarray}}
\def\eeq{\end{eqnarray}}

\title{Perpendicular Reading of Single Confined Magnetic Skyrmions}
\author{Dax M. Crum}
\affiliation{Microelectronics Research Center, The University of Texas at Austin \\10100 Burnet Road, Austin, Texas 78758, USA and\\	
			Peter Gr{\"u}nberg Institut and Institute for Advanced Simulation\\Forschungszentrum J{\"u}lich and JARA, D-52425 J{\"u}lich, Germany}
\email{dcrum@utexas.edu}
\author{Mohammed Bouhassoune}
\author{Juba Bouaziz}
\author{Benedikt Schweflinghaus}
\author{Stefan Bl{\"u}gel}
\author{Samir Lounis}
\affiliation{Peter Gr{\"u}nberg Institut and Institute for Advanced Simulation\\Forschungszentrum J{\"u}lich and JARA, D-52425 J{\"u}lich, Germany}
\date{\today}

\begin{abstract}
{\fontfamily{phv}\selectfont Thin-film sub-5 nm magnetic skyrmions constitute an ultimate scaling alternative for future digital data storage. Skyrmions are robust non-collinear spin-textures that can be moved and manipulated by small electrical currents. We show here an innovative technique to detect isolated nano-skyrmions with a current-perpendicular-to-plane geometry, which has immediate implications for device concepts. We explore the physics behind such a mechanism by studying the atomistic electronic structure of the magnetic quasiparticles. We investigate how the isolated skyrmion local-density-of-states which tunnels into the vacuum, when compared to the ferromagnetic background, is modified by the site-dependent spin-mixing of electronic states with different relative canting angles. Local transport properties are sensitive to this effect, as we report an atomistic conductance anisotropy of up to $\sim$20\% for magnetic skyrmions in Pd/Fe/Ir(111) thin-films. In single skyrmions, engineering this spin-mixing magnetoresistance possibly could be incorporated in future magnetic storage technologies. }
\end{abstract}
\maketitle
\section*{ }
Si complimentary-metal-oxide-semiconductor$^{1}$ (CMOS) compatible magnetic devices represent the current state-of-the-art in information data storage circuits$^{2}$. In such devices, the information is encoded by manipulation of different spacial magnetic domains, and the data is read by sensing the variation in the magnetoresistance as a function of the magnetization direction$^{3}$.

Incorporation of magnetic skyrmions instead of domain walls (DWs) may improve device performance and scaling possibilities$^{4}$. DWs are sensitive to defect-pinning$^{5-7}$ while skyrmions intrinsically are not$^{8-10}$. This insensitivity to defects explains why skyrmions can be moved by electric current densities orders of magnitude smaller than DWs$^{11,12}$ while achieving smaller yet still sizeable magnitudes of velocity$^{10,13-15}$. Skyrmions, for topological reasons$^{16}$, are relatively robust magnetic particle-like configurations$^{17,18}$, stable even up to, so far, near-room temperature$^{19}$. They can be confined at will and their shape and size controlled with an external magnetic field$^{20}$. 

An illustrative device scaling route is an extension$^{4}$ of the racetrack memory$^{21}$ to incorporate single$^{22,23}$ confined magnetic skyrmions instead of DWs. Such a concept constitutes an ultimate scaling alternative in terms of packing density, speed, and power consumption$^{4}$. We envision a thin-film magnetic heterostructure$^{24}$ where sub-5 nm chiral skyrmionic quasiparticles are generated (via materials engineering and external magnetic fields) and moved laterally along a magnetic racetrack by in-plane currents, which has been shown experimentally$^{11,12,25}$. The bit-wise data would be encoded by out-of-plane currents which create or annihilate the individual skyrmions, thus setting or resetting the bit-state. This has also been shown experimentally$^{23}$; thus, two important ingredients to a viable skyrmion racetrack memory system, lateral bit-wise movement and set/reset of each bit-state, have been established experimentally.

But how could one \emph{read} each bit-state? Current-in-plane (CIP) detection of skyrmions has been shown experimentally$^{25-27}$ and understood theoretically as a topological Hall effect$^{28,29}$, but may be costly in terms of power consumption and difficult to fabricate in terms of device geometries. A better option would be the direct detection of the nano-skyrmions via current-perpendicular-to-plane (CPP)$^{30,31}$ geometry.

\begin{figure*}[t!]
        \includegraphics[width=0.95\textwidth]{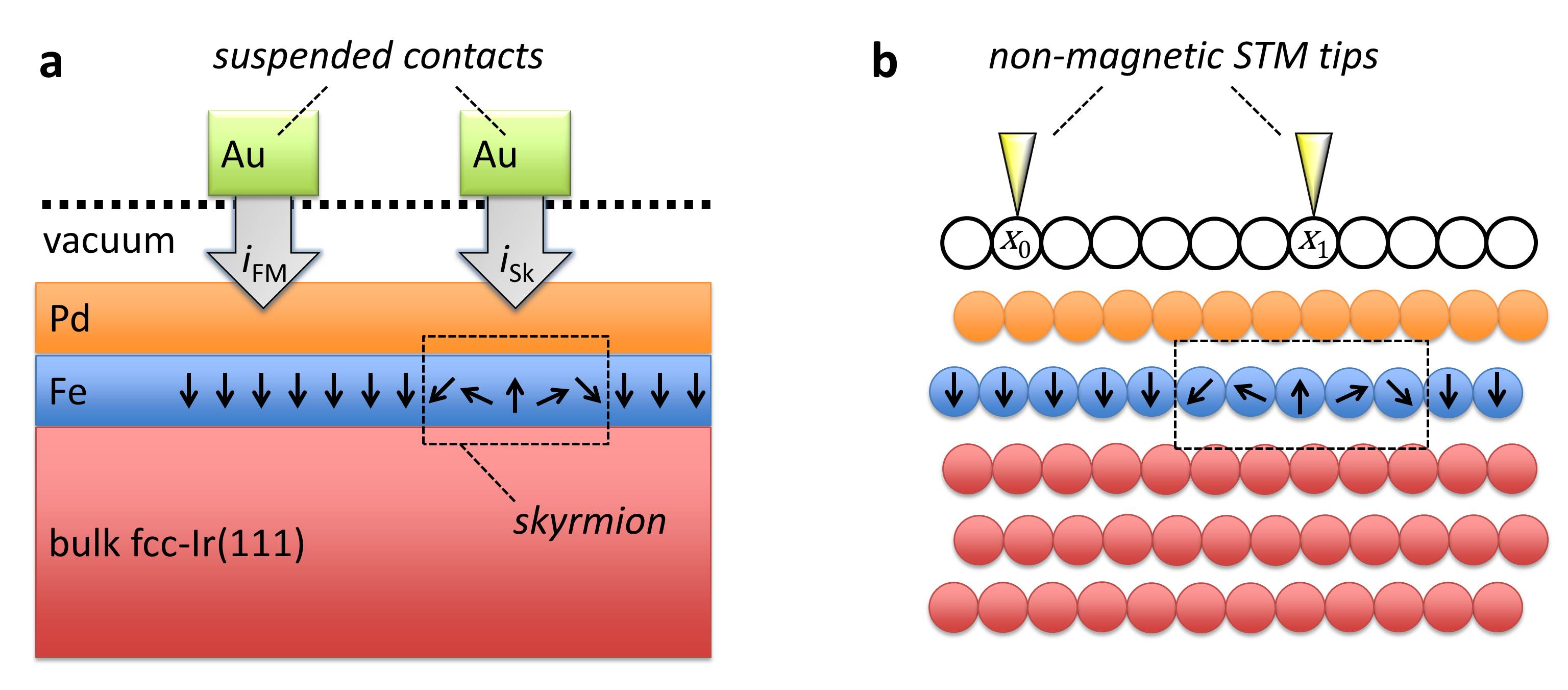}
	\caption{\label{fig:mat_stack} \textbf{\textbar\; Perpendicular detection of nano-skrymions. (a)} Illustrative concept device for perpendicular reading of single nano-skyrmions. Due to energy-dependent spin-mixing perturbations in the atomistic electronic structure as a function of position within skyrmions, the electric current relation $i_{\mathrm{FM}}\neq i_{\mathrm{Sk}}$ holds, and therefore magnetic data information can be sensed in a CPP-geometry. \textbf{(b)} Illustrative STM-spectroscopy experiment of fcc-Pd/Fe overlayer on single-crystal fcc-Ir(111) bulk substrate. The tunneling conductance is modified by the combined effects of local magnetic non-collinearity and substrate-induced spin-orbit interaction. For similar physical reasons as (\textbf{a}), the tunneling conductance at position $x_0$ is different from that at position $x_1$. }
\end{figure*}

We demonstrate that such a detection mechanism for single skyrmions is indeed possible, even using non-spin-polarized electric currents. Using Fig.~\ref{fig:mat_stack}a as a schematic illustration, we will show how the tunneling current between a suspended metal contact through vacuum depends on the non-collinear magnetic state-of-phase below. (Suspended metal contacts are possible with state-of-the-art fabrication techniques$^{32}$, but one could \emph{also} imagine tunneling through a weakly-interacting two-dimensional (2D) insulator, such as hBN or MoS$_2$). Such a process can be intimately understood in a non-spin-polarized scanning tunneling microscopy (STM) experiment, Fig.~\ref{fig:mat_stack}b. This all-electrical detection is a departure from typical experiments which rely on spin-polarized injection to detect magnetic structures, and thus inherently an improvement from a device application perspective. 

We will examine how the tunneling current between a non-magnetic STM tip and surface varies as a function of the skyrmion's local atomic magnetization direction. Considering the skyrmion's central spin-flipped atom, its local magnetic environment is inequivalent from those atoms in the ferromagnetic background. Thus, one could expect already a conductance anisotropy in the center of the skyrmion due to the change in the electronic structure relative to the ferromagnetic state. One purpose of our work here is to investigate this effect. In addition, the heavy-metal substrate induces a large spin-orbit interaction (SOI), coupling the local magnetization to the real-space direction, further modifying the electronic structure as a function of magnetization rotation relative to the substrate plane. This tunneling anisotropic magnetoresistance (TAMR)$^{33}$ effect has been studied in detail but typically for homogeneous magnetic domains (either different ferromagnetic domains$^{33,34}$ or homogeneous spin-spirals$^{35}$). We seek to extend this theoretical basis to inhomogeneous spin-textures such as skyrmions. We focus specifically on \emph{single} skyrmions and do not investigate networks or lattices of skyrmions.

In this Article, we explain the combined (non-collinear and SOI-induced) spin-mixing magnetoresistance in terms of fully self-consistent calculations from first-principles, where we have direct access to the electronic structure of not just the magnetic heterostructure, but even of the states decaying into the vacuum, which are essential to the tunneling conductance. We find a rather large atomistic conductance anisotropy of up to $\sim$20\% ($\sim$10\%) for magnetic skyrmions in Pd/Fe/Ir (Pd/Pd/Fe/Ir) magnetic thin-films, which potentially could be detected in a CPP-device geometry. Establishing the physics of this generalized tunneling spin-mixing magnetoresistance (TXMR) could possibly inspire the design of future nano-magnetic devices based on such a mechanism.
\section*{Results}
\subsection*{System and procedure}
We study two magnetic thin-film heterostructures similar to Fig.~\ref{fig:mat_stack}b purely from \emph{ab initio}: fcc-overlayers of Pd/Fe and Pd/Pd/Fe on single-crystal bulk fcc-Ir(111). These systems are attractive for a number of reasons. First, they generate large Dzyaloshinskii-Moriya interactions (DMIs)$^{36-38}$, whose competition with the isotropic exchange interaction $J$ determines the size and chirality of the skyrmions$^{4}$. DMIs are large here because of the strength and nature of the inversion-symmetry-breaking in the heterostructures. At the Fe/Ir(111) interface, a large spin-orbit interaction in the underlying heavy-metal substrate, here Ir(111), is relatively uncompensated by the overlayer Pd/Fe or Pd/Pd/Fe interface, leading to a large DMI vector preferentially in the plane of Fe, denoted by $\mathbf{D}$. The ratio of $|\mathbf{D}|/J$, along with an external magnetic field, can stabilize isolated skyrmions with diameter $D_\mathrm{{Sk}}\approx$1-5 nm in size, and has been shown experimentally$^{20,23}$. Second, by choosing a double-Pd overlayer (Pd/Pd/Fe/Ir) versus a single-Pd overlayer (Pd/Fe/Ir), one can alter the exchange interactions in Fe due to the modified nature of the interface hybridization and electronic charge transfer (see Supplementary Information S1). We investigate this effect to illuminate conceptual studies where other overlayer combinations and materials are used to engineer the size, shape, and stability of the isolated skyrmions$^{24,39,40}$. 

We perform self-consistent density functional theory (DFT) calculations based on a full-potential Green function formalism including SOI$^{41,42}$, which allows a perfect embedding of \emph{real}-space defects, such as isolated skyrmions, into the ferromagnetic background system. Additional specifics of our computational scheme are given in the Methods section. 

\subsection*{Non-collinear inhomogeneity in nano-skyrmions}
\begin{figure*}[t!]
	\includegraphics[width=0.95\textwidth]{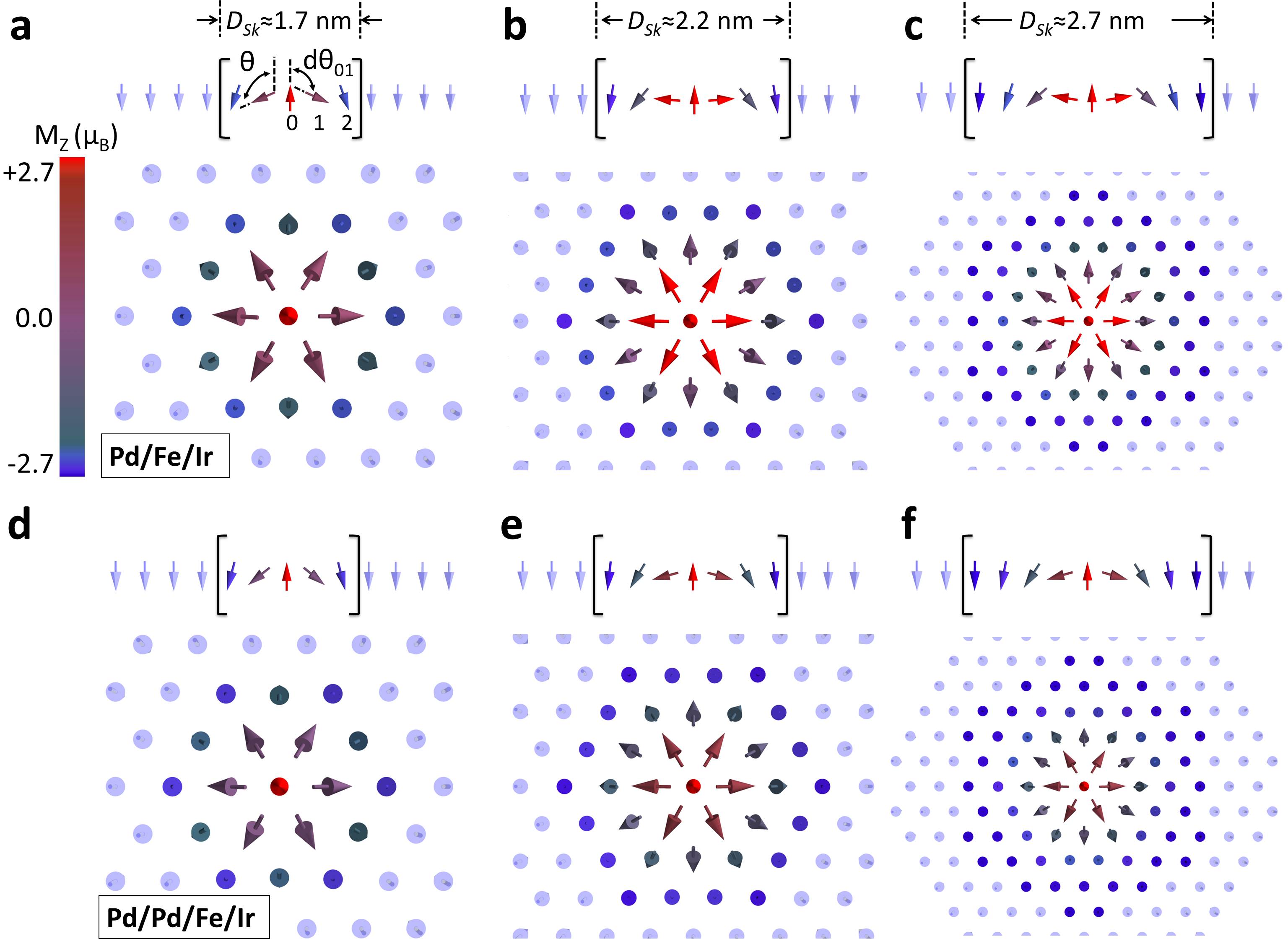}
	\caption{\label{fig:all_spins} \textbf{\textbar\; Real-space relaxation of nano-skyrmions with increasing size. (a-c)} Plots of axisymmetric cycloidal spin-whirls inside a magnetically active Fe-layer centered about increasingly larger skyrmionic defects in fcc-Pd/Fe overlayer on fcc-Ir(111) bulk substrate. Confining spins in the FM-background are shown transparent. We define $\theta$ as the typical polar angle with the vertical and d$\theta$ as the difference in polar angle between adjacent pairwise atoms. \textbf{(d-f)} Again but in fcc-Pd/Pd/Fe overlayer on fcc-Ir(111) bulk substrate. The smallest defects we study ($D_{\mathrm{Sk}}\approx 1.1$~nm) are not shown here. The color bar in (\textbf{a}) represents for all illustrations.}
\end{figure*}

Before coming to the essential physics of the tunneling spin-mixing magnetoresistance (TXMR) effect, we first self-consistently relax different-sized nano-skyrmions in otherwise ferromagnetic backgrounds (see Fig.~\ref{fig:all_spins}), in both single- and double-Pd overlayer material stacks. We control the size of the skyrmionic defects by allowing different finite numbers of atoms to relax their magnetic moments in size and direction after the central atom has been spin-flipped as an initial condition.  We investigate four different realistic skyrmion sizes: $D_{\mathrm{Sk}} \approx$ 1.1, 1.7, 2.2, and 2.7~nm in diameter. The spin-textures exhibit a fixed and unique rotational sense as demanded by the DMI, which seeks energy gain by torquing the moments to rotate with respect to their neighbors. These structures are cycloidal and radial in nature as expected for magnetic thin-films. Thus our theoretical calculations are consistent in generating realistic nano-skyrmions which have been experimentally detected using \emph{magnetic} spin-polarized currents$^{20,23}$.

We illustrate the spin-moment global rotation vs.\ the vertical (polar angle $\theta$) of each atom and the pairwise difference between adjacent polar angles (d$\theta$). We will show that the spin-mixing perturbations to the local-density-of-states (LDOS)  are a function of these angular parameters because the relative canting between different pairwise atomic sites varies as a function of space, in addition to the absolute canting relative to the substrate. When analyzing each individual confined structure, we mention that d$\theta$ itself is not constant between different pairs of atoms, such that there exists an inhomogeneity on the atomic-scale in the rotation of the magnetization direction with respect to the substrate plane. Furthermore, these inhomogeneities themselves are a function of diameter when comparing skyrmions of different sizes (Fig.~\ref{fig:all_spins}a,b,c). 

\subsection*{Electronic structure of isolated confined skyrmions}
We now move to establish the physics behind the TXMR effect within a scanning tunneling microscopy/spectroscopy (STM or STS) experiment employing a \textit{non}-spin-polarized tip, for which according to the Tersoff-Hamann model$^{43}$, the differential conductance d$I$/d$V$ is proportional to the LDOS of the sample, calculated at the tip position, $\mathbf{R}_{\mathrm{tip}}$, and the given bias energy $E_{\mathrm{bias}}$:
\begin{equation}
 \frac{\mathrm{d}I}{\mathrm{d}V} \propto \mathrm{LDOS}\,(\,\mathbf{R}_{\mathrm{tip}}; E_{\mathrm{bias}}, \{\mathbf{s}_i\}) \;\;.
\end{equation}
The LDOS depends on the configuration $\{\mathbf{s}_i\}$ of atomic spins of the sample (i) relative to each other, e.g.\ in terms of d$\theta_{ij}=\theta_j-\theta_i$ for all atom-pairs $(i,j)$, and (ii) relative to the lattice in terms of the absolute polar angle $\theta_i$. The transport phenomenon related to the latter is known as the tunneling anisotropic magnetoresistance (TAMR)$^{33}$, an effect related to spin-mixing due to the SOI. The former results from the spin-mixing hybridization of majority- and minority-states due to non-collinearity. Both can be subsumed  as tunneling spin-mixing magnetoresistance (TXMR). Common to both is that the probability of tunneling into majority- and minority-states depends on angles. The difference is that (i) is expected to be larger than (ii), since it is not a SOI effect. Also the appearance of both are different. For example, the TXMR due to (i) in a homogeneous magnetic spiral is the same across the spiral, because d$\theta_{ij}=$\;d$\theta$ for all atom-pairs $(i, j)$, but different for spirals of different pitches under the transformation d$\theta\rightarrow\;$d$\theta^\prime$. In contrast, the TAMR is modulated \emph{across} the spiral$^{35}$ as $\theta_i$ changes from atom-to-atom.  Thus, the TXMR is used to measure conductance differences between two different magnetic states such as the difference between a skyrmion and the FM-state, but can also be used to resolve magnetization inhomogenieties inside complex spin-textures such as skyrmions or domain-walls. 

\begin{figure*}[t!]
        \includegraphics[width=0.95\textwidth]{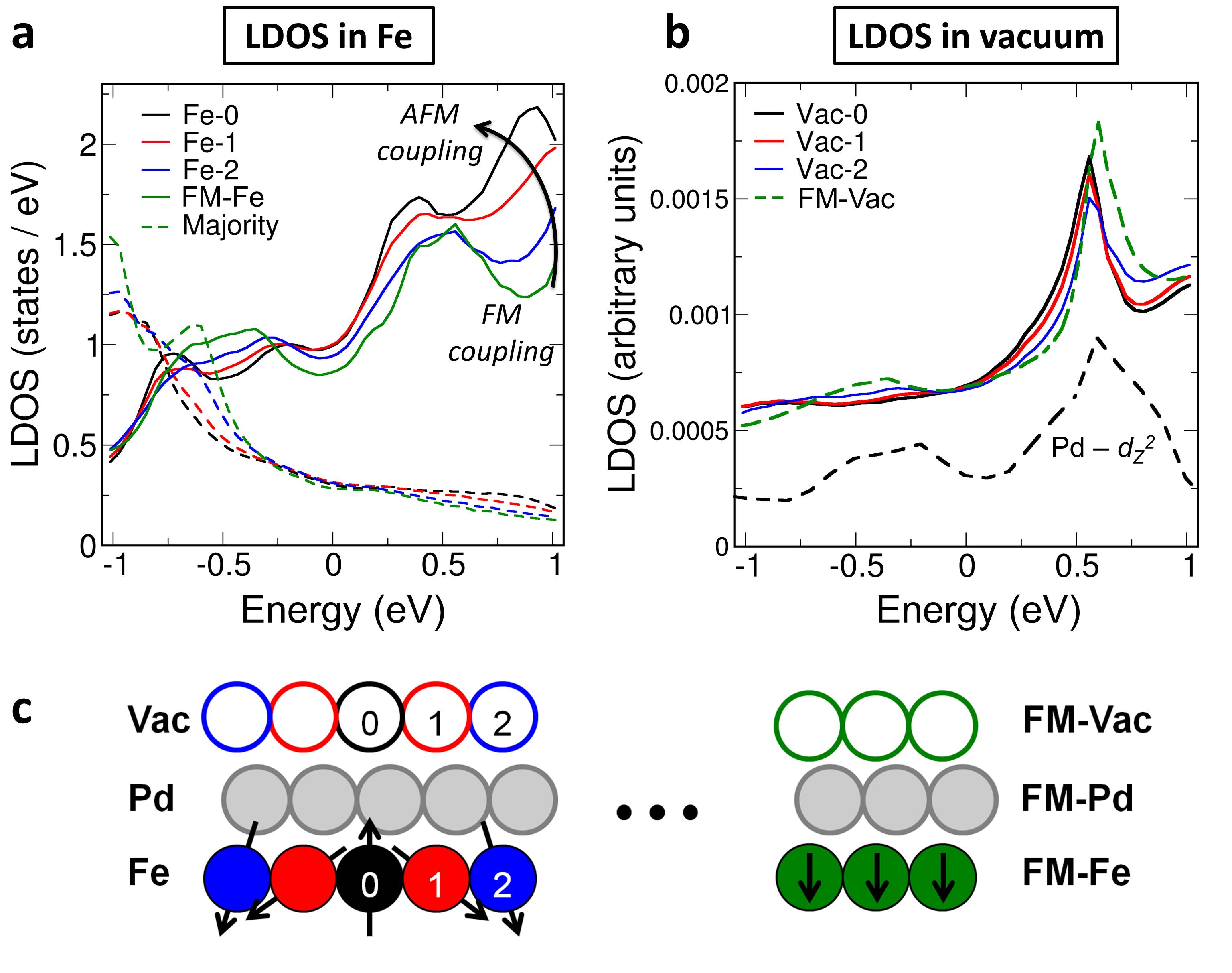}
	\caption{\label{fig:ldoss} \textbf{\textbar\; LDOS in a realistic nano-skyrmion in Pd/Fe/Ir. (a)} LDOS in the magnetically-active Fe-layer resolved into minority and majority spin-channels. The resonance-peak near $E\approx 0.5$ eV in the FM-background (green) shifts in energy when approaching the center of the skyrmion (black). \textbf{(b)} The modification of the electronic structure in Fe contributes to a strong resonance in the LDOS in vacuum via hybridization through surface Pd-states. Arbitrary units are used so as to include in the same plot the nature of the Pd-$d_{z^2}$ surface-state, whose resonance-peak near 0.5~eV survives in the vacuum. \textbf{(c)} Illustrative legend for (\textbf{a}, \textbf{b}) where the atoms are numbered and color-coded to identify the appropriate curves to the corresponding atoms in the skyrmion. The vacuum domains are represented by empty spheres. }
\end{figure*}
Since the spin-mixing perturbations due to non-collinearity and SOI are magnetic in nature, we show in Fig.~\ref{fig:ldoss}a the spin-dependent LDOS in the magnetically-active Fe-layer as a function of the atomic position for the $D_{\mathrm{Sk}} \approx 1.7$~nm skrymion. For brevity we plot only the Pd/Fe/Ir case (see Supplementary Information S2 for the larger $D_{\mathrm{Sk}} \approx 2.2$~nm case and S3 for the double-Pd cases). We note that the majority and minority spin-channels are given in the local spin-frame of reference for each Fe-atom. The color-coding of the plot, which corresponds to different atoms extending radially from the skyrmion's center, is explained in Fig.~\ref{fig:ldoss}c. The energy-zero is the Fermi energy, $E_F=0$. 

The resonant-states between $0.5$ and $1.0$~eV above $E_F$ are of Fe $d$-band minority-spin character and consist of $d_{xy/x^2-y^2}$, $d_{xz/yz}$, and $d_{z^2}$ states. These states hybridize with the $sp$ states in the Pd-overlayer and give rise to Fe-Pd-$spd_{z^2}$ hybrid states, named in short Pd-$d_{z^2}$ states localized in the Pd overlayer-film around $E\approx$\;0.5 eV, as shown in Fig.~\ref{fig:ldoss}b (black-dashed curve). It is clear that the surface-layer Pd-$d_{z^2}$ state (shown only for the background-FM Pd-surface film), which has the proper orbital symmetry to decay slowly transverse to the substrate, controls the electronic structure in vacuum as a function of energy, characterized by a strong resonance in the vacuum-LDOS. An all-electrical STS measurement will be sensitive to this vacuum resonance-peak.

The origin of this resonance and its behavior upon rotation can be understood by analyzing the energy window near $E\approx0.5$~eV in Fig.~\ref{fig:ldoss}a, where one can see the resonance-peaks shifting in energy in the Fe skyrmion-LDOS as a function of position. The green curve, which represents the ferromagnetic state of the background Fe-film, shows an electronic structure consistent with Fe-minority $d$-$d$ hybridization when adjacent atoms couple ferromagnetically (see Supplementary Section S2). Conversely, moving towards the center of the skyrmion, the quantization axes between two neighboring atoms becomes different, and majority-states of one atom can hybridize with minority-states of the second. This effect is especially pronounced at the central spin-flipped atom (black curve), where the resonance-peak has shifted lower in energy \--- as expected for \emph{antiferromagnetic}  (AFM) coupling (see Supplementary Section S2). 
We reproduce these effects within the context of a simple model, where we can qualitatively predict the change in LDOS as a function of the non-collinear magnetization rotation parameter d$\theta$ as defined in Fig.~\ref{fig:all_spins}a (see Supplementary Information S4 and S5).

The energy-dependent disturbance to the LDOS resonance-peaks as a function of position moving radially along the skyrmion will manifest as a perturbation to the local electrical conductivity, and is the physical basis for the space-dependent TXMR effect.

\subsection*{All-electrical skyrmion detection}
We now define the TXMR and make predictions for future experimental observation of the effect. The TXMR is the percent deviation of the local conductance from a reference conductance due to the spin-mixing from non-collinearity and SOI. As long as the magnetic state under consideration has a different non-collinearity than the reference state, there will be a TXMR. If one is interested in the spatial resolution of a complex spin texture  (ignoring SOI), however, then an additional inhomogeneity within the non-collinearity is required, as is the case for nano-skyrmions.

The TXMR is by definition measured in vacuum. Here, we choose the reference to be somewhere far from the skyrmion in the FM-background. Then the normalized TXMR measured at site $r$ is
\begin{equation}
\mathrm{TXMR}(r) = \frac{\mathrm{LDOS}_{\mathrm{FM}}^{\mathrm{vac}}-\mathrm{LDOS}_{\{\mathbf{S}\}}^\mathrm{vac}(r)}{\mathrm{LDOS}_{\mathrm{FM}}^{\mathrm{vac}}}\times 100\% \;\; ,
\end{equation}
where $\mathrm{LDOS}_{\mathrm{FM}}^{\mathrm{vac}}$ is the LDOS in the vacuum just above the FM, and $\mathrm{LDOS}_{\{\mathbf{S}\}}^\mathrm{vac}(r)$ is the LDOS of the complex spin-texture in the vacuum just above site $r$.

Integrating the TXMR over the entire device injection boundary, over all energies up to the bias energy $eV_{\mathrm{bias}}$, would give a measure of the total change in conductance, and would be the state-of-bit detection mechanism in a CPP-TXMR device like discussed in Fig.~\ref{fig:mat_stack}a. In an STS experiment, however, the effect could be amplified by selecting specific energy windows where the TXMR were largest as a function of position.

\begin{figure*}
	\includegraphics[width=0.90\textwidth]{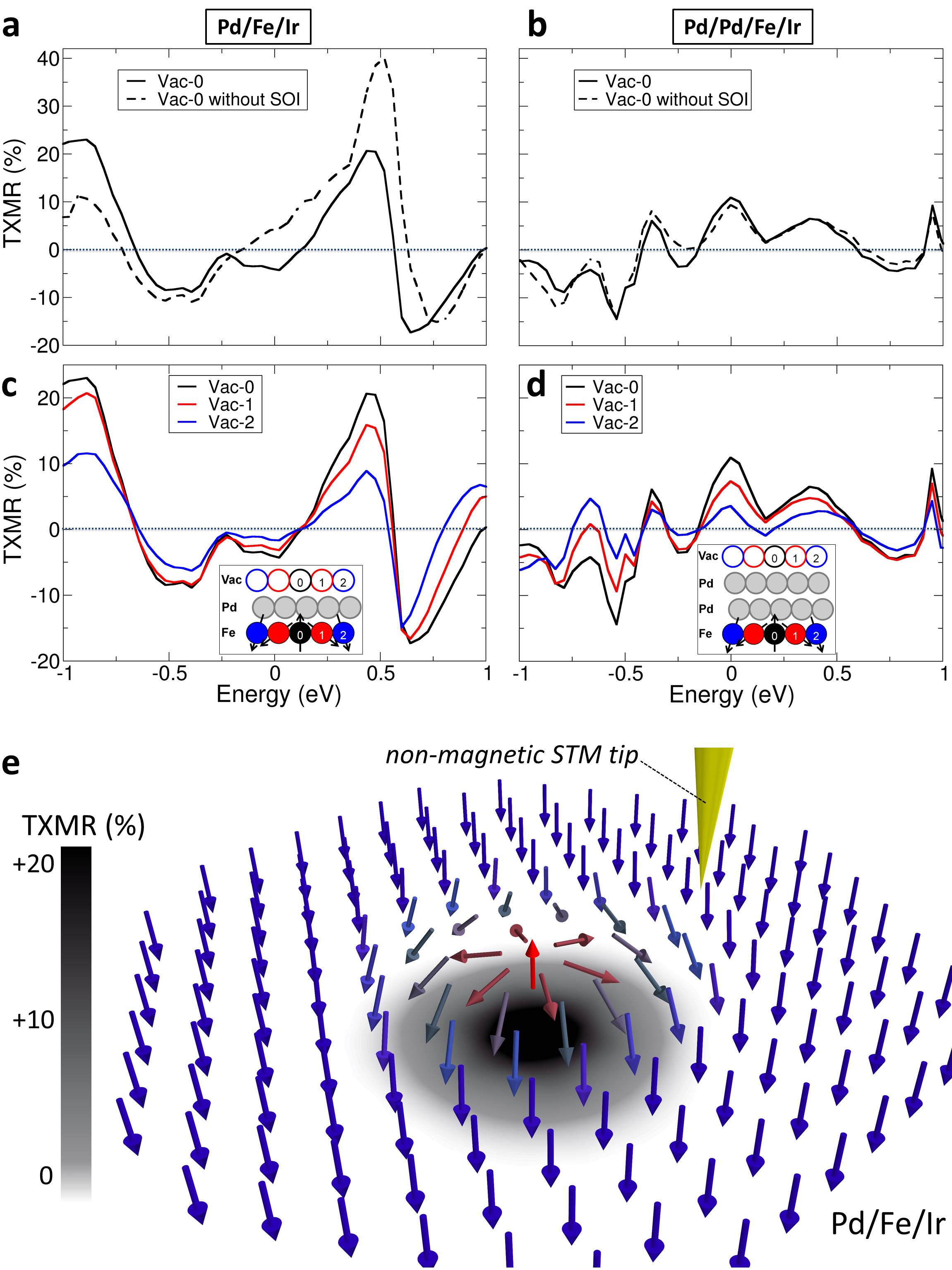}
	\caption{\label{fig:txmr} \textbf{\textbar\; Tunneling spin-mixing magnetoresistance. (a,b)} TXMR signals comparing the effects of SOI in a $D_{\mathrm{Sk}} \approx$ 1.7 nm skyrmion, for single- and double-Pd systems. \textbf{(c,d)} Again for a line of atoms extending radially (see insets for labeling convention). \textbf{(e)} Expected STS-signal when approaching a skyrmionic defect in the single-Pd case. The electrical contrast has been projected onto the plane below the skyrmion.}
\end{figure*}

In Fig.~\ref{fig:txmr}a,b we show the energy-resolved TXMR of the $D_{\mathrm{Sk}} \approx$ 1.7 nm skyrmion's central spin-flipped vacuum-site, with and without SOI, for the single- and double-Pd cases. We notice a sizeable TXMR effect for both systems. This holds true for all skyrmions that we studied, noting a small size-dependence of the effect which varies weakly as a function of $D_{\mathrm{Sk}}$ (see Supplementary Information S6). 

TXMR signals in the different single-Pd and double-Pd material systems vary from skyrmion-to-skyrmion, however. Since the SOI arises from the Fe-Ir interface, the impact of SOI is much more pronounced in the single-Pd system (Fig.~\ref{fig:txmr}a), where the TXMR can peak out at an impressive $\sim$40\% when ignoring spin-orbit coupling, but decreases down to $\sim20$\% when SOI is included. Interestingly, the spin-mixing due to SOI is to compete with the effects due to inhomogeneous non-collinearity, reducing the overall TXMR signal. In the double-Pd system, most of the TXMR signal is due to inhomogeneous non-collinearity, with a small contribution coming from the SOI (Fig.~\ref{fig:txmr}b).

In Fig.~\ref{fig:txmr}c,d we plot the spacial variation of the TXMR signal, which differs significantly from atom-to-atom within the same skyrmion. We see that the TXMR effect is reduced when approaching the edge of the skyrmion (Fig.~\ref{fig:txmr}c,d blue curves), since the effective non-collinearity is reduced as the complex spin-texture fades into the ferromagnetic background.

The vacuum-resonance we found in Fig.~\ref{fig:ldoss}b appears now as a large TXMR signal at the same energy in the Pd/Fe/Ir(111) system (Fig.~\ref{fig:txmr}a,c). Thus, an experimentalist probing the surface in a STS experiment, having set the ripple-bias voltage near $V_{\mathrm{bias}}\approx 0.5$V, would see an electrical contrast as visualized in Fig.~\ref{fig:txmr}e when approaching a skyrmion of similar size in the single-Pd heterostructure.

Within a reasonable bias voltage range, the TXMR effect is smaller in the double-Pd case ($\sim$10\%) when compared to the single-Pd case ($\sim$20\%). The additional Pd-overlayer changes the resonance nature around 0.5 eV compared to the single-Pd case, and states with a high tunneling cross-section into the vacuum are distributed over a wider energy (see Supplementary Information S3). As a consequence, the TXMR is reduced by nearly half when compared to the single-Pd case, and a bias energy near $-0.5$~eV is experimentally more favorable. 
\section*{Discussion}
We have studied realistic- and experimentally-observable confined nano-skyrmions within metallic thin-films of Pd/Fe/Ir(111) and Pd/Pd/Fe/Ir(111) completely from first-principles. We established how the combined effects of local inhomogeneous magnetic non-collinearity and SOI in nano-skyrmions can alter the atomistic electronic structure in a magnetically active Fe-film, and, via hybridization with additional surface layers, the electrons which tunnel into the vacuum. 

The change in the LDOS can be understood in terms of the rotation parameters of the magnetic moment of the considered atom. The largest spin-mixing contribution comes from non-collinearity and depends on the relative canting between magnetic moments on neighboring sites, d$\theta$. The dependence on the absolute polar angle of the magnetic moment with respect to the substrate, $\theta$, comes in as a second-order term to the change in the LDOS, but can become important if the impact of the SOI is large. 

Finally, we have shown in detail how such a physical interplay could induce a sizeable electrical conduction anisotropy as a function of position and energy in realistic nano-skyrmions, up to $\sim$20\% in the single-Pd case.  The manifestation of this TXMR effect could possibly be exploited in an all-electrical tunneling spectroscopy experiment. 

In addition, the changes in the magnetoresistance on the nanometer scale of skyrmions can possibly be engineered to design advanced magnetic memory devices. Typical memory circuits require at least one control device (either transistor or diode) in each memory cell. Instead, technologies based on spin-mixing in single skyrmions could have potentially hundreds of bits stored in nanometer-sized magnetic racetracks needing only a single read-out element to detect the contents of each array (see Supplementary Information S7).

\section*{Methods}
The electronic structure was determined employing density functional theory (DFT) in the local spin density approximation$^{44}$. Calculations were executed by means of the screened Korringa-Kohn-Rostoker full-potential scalar-relativistic Green function method$^{41,42}$, where spin-orbit coupling was included self-consistently. A full-potential method is important to accurately describe the nature of the complex spin-texture and rapidly-decaying vacuum states of the tunneling electrons. 

For the calculations we chose an angular momentum cutoff of $l_{max}$ = 3 for the orbital expansions of the Green functions. The energy contour for numeric integration of the spin and charge density contained 40 grid points in the upper complex plane (including 7 Matsubara poles) with a Brillouin zone mesh of 30$\times$30 $k$-points. The FM-slab LDOS and skyrmion impurity cluster LDOS were obtained by one-shot calculations using the FM-state or skyrmion-state as starting points, respectively. We found that increasing the $k$-mesh to 200$\times$200 was sufficiently adequate to numerically stabilize the relevant observables.

The magnetic thin-film slab configurations follow, where positive percentages refer to inward relaxation with respect to the Ir(111) interlayer separation. We consider only fcc-stacking in all cases, which is in fact energetically favorable compared to hcp crystal growth$^{39}$. 

(a) Pd/Fe/Ir: 44 total layers (3 vacuum + 1 vacuum (-1\%) + 1 Pd (8\%) + 1 Fe (7\%) + 1 Ir (1\%) + 33 Ir + 4 vacuum). 

(b) Pd/Pd/Fe/Ir: 44 total layers (3 vacuum + 1 Pd (-1\%) + 1 Pd (8\%) + 1 Fe (7\%) + 1 Ir (1\%) + 33 Ir + 4 vacuum). 

We choose 34 Ir layers since it was the minimum thickness by which we completely decoupled any wave function penetration from top-to-bottom surface. We obtained the relaxation parameters as optimized and reported by Dup{\'e} \emph{et al}$^{39}$.

To stabilize skyrmions after determining the 2D-FM slabs, the slab Green functions were harvested and a single spin-flipped Fe-atom was embedded in the FM-background. We then allowed three-layer cylindrical ring-like stacks of atoms within the skyrmion impurity cluster to update their potentials and magnetic moments (Fe-layer + 1 Pd-layer above and 1 Ir-layer below). The effect of the FM-background was included self-consistently by the slab Green function ($G_0$), which connects the skyrmion impurity cluster ($G_{imp}$) to the host via the Dyson-like equation: $G_{imp} = G_0 + G_0\,\Delta V\, G_{imp}$, where $\Delta V$ represents the modified atomic potential as compared to the unperturbed slab Green function potential, $V$. In such a manner a real-space defect can be perfectly embedded in an otherwise periodic crystal. After converging the different sized skyrmionic profiles, observables were calculated as mentioned previously.
\section*{References}
1. Sze, S. M. \& Ng, K. K. \emph{Physics of Semiconductor Devices}, 3rd ed. (Wiley, Hoboken, New Jersey, 2007).

2. Chappert, C., Fert, A. \& Dau, F. N. V. The emergence of spin electronics in data storage. \emph{Nature Mater.} \textbf{6}, 813 (2007).

3. Thompson, D. A. \& Best, J. S. The future of magnetic data storage techology. \emph{IBM Journal of Research and Development} \textbf{44}, 311 (2000).

4. Fert, A., Cros, V. \& Sampaio, J. Skyrmions on the track. \emph{Nature Nanotech.} \textbf{8}, 152 (2013).

5. Kl{\"a}ui, M. \emph{et al}. Direct observation of domain-wall pinning at nanoscale constrictions. \emph{Appl. Phys. Lett.} \textbf{87}, 102509 (2005).

6. Martinez, E., Lopez-Diaz, L., Alejos, O., Torres, L. \& Tristan, C. Thermal effects on domain wall depinning from a single notch. \emph{Phys. Rev. Lett.} \textbf{98}, 267202 (2007).

7. Jang, Y. \emph{et al}. Current-induced domain wall nucleation and its pinning characteristics at a notch in a spin-valve nanowire. \emph{Nanotechnology} \textbf{20}, 125401 (2009).

8. Iwasaki, J., Mochizuki, M. \& Nagaosa, N. Universal current-velocity relation of skyrmion motion in chiral magnets. \emph{Nature Commun.} \textbf{4}, 1463 (2013).

9. Iwasaki, J., Mochizuki, M. \& Nagaosa, N. Current-induced skyrmion dynamics in constricted geometries. \emph{Nature Nanotech.} \textbf{8}, 742 (2013).

10. Sampaio, J., Cros, V., Rohart, S., Thiaville, A. \& Fert, A. Nucleation, stability, and current-induced motion of isolated magnetic skyrmions in nanostructures. \emph{Nature Nanotech.} \textbf{8}, 839 (2013).

11. Jonietz, F. \emph{et al}. Spin transfer torques in MnSi at ultralow current densities. \emph{Science} \textbf{330}, 1648 (2010).

12. Yu, X. Z. \emph{et al}. Skyrmion flow near room temperature in an ultralow current density. \emph{Nature Commun.} \textbf{3}, 988 (2012).

13. Thiaville, A., Nakatani, Y., Miltat, J. \& Suzuki, Y. Micromagnetic understanding of current-driven domain wall motion in patterned nanowires. \emph{Europhys. Lett.} \textbf{69}, 990 (2005).

14. Khalkovskiy, A. V. \emph{et al}. Matching domain-wall configuration and spin-orbit torques for efficient domain-wall motion. \emph{Phys. Rev. B} \textbf{87}, 020402 (2013).

15. Khalkovskiy, A. V. \emph{et al}. High domain wall velocities due to spin currents perpendicular to the plane. \emph{Phys. Rev. Lett.} \textbf{102}, 067206 (2009).

16. Skyrme, T. H. R. A unified field theory of mesons and baryons. \emph{Nucl. Phys.} \textbf{31}, 556 (1962).

17. R{\"o}{\ss}ler, U. K., Bogdanov, A. N. \& Pfleiderer, C. Spontaneous skyrmion ground states in magnetic metals. \emph{Nature} \textbf{442}, 791 (2006).

18. Yu, X. Z. \emph{et al}. Real-space observation of a two-dimensional skyrmion crystal. \emph{Nature} \textbf{465}, 901 (2010).

19. Yu, X. Z. \emph{et al}. Near room-temperature formation of a skyrmion crystal in thin-films of the helimagnet FeGe. \emph{Nature Mater.} \textbf{10}, 106 (2011).

20. Romming, N., Kubetzka, A., Hanneken, C., von Bergmann, K. \& Wiesendanger, R. Field-dependent size and shape of single magnetic skyrmions. \emph{Phys. Rev. Lett.} \textbf{114}, 177203 (2015).

21. Parkin, S. S. P., Hayashi, M. \& Thomas, L. Magnetic domain-wall racetrack memory. \emph{Science} \textbf{320}, 190 (2008).

22. Kiselev, N. S., Bogdanov, A. N., Sch{\"a}fer, R. \& R{\"o}{\ss}ler, U. K. Chiral skyrmions in thin magnetic films: new objects for magnetic storage technologies? \emph{J. Phys. D} \textbf{44}, 392001 (2011).

23. Romming, N. \emph{et al}. Writing and deleting single magnetic skyrmions. \emph{Science} \textbf{341}, 636 (2013).

24. Dup{\'e}, B. , Bihlmayer, G., Bl{\"u}gel, S. \& Heinze, S. Engineering skyrmions in transition-metal multilayers for spintronics. arXiv:1503.08098 (2015).

25. Schulz, T. \emph{et al}. Emergent electrodynamics of skyrmions in a chiral magnet. \emph{Nature Phys.} \textbf{8}, 301 (2012).

26. Neubauer, A. \emph{et al}. Topological Hall effect in the A Phase of MnSi. \emph{Phys. Rev. Lett.} \textbf{102}, 186602 (2009).

27. Kanazawa, N. \emph{et al}. Discretized topological Hall effect emerging from skyrmions in constricted geometry. \emph{Phys. Rev. B} \textbf{91}, 041122 (2015).

28. Freimuth, F., Bamler, R., Mokrousov, Y. \& Rosch, A. Phase-space Berry phases in chiral magnets: Dzyaloshinskii-Moriya interaction and the charge of skyrmions. \emph{Phys. Rev. B} \textbf{88}, 214409 (2013).

29. Franz, C. \emph{et al}. Real-space and reciprocal-space Berry phases in the Hall effect of Mn$_{1\mathrm{-}x}$Fe$_x$Si. \emph{Phys. Rev. Lett.} \textbf{112}, 186601 (2014).

30. Pratt, W. P. \emph{et al}. Perpendicular Giant Magnetoresistances of Ag/Co Multilayers. \emph{Phys. Rev. Lett.} \textbf{66}, 3060 (1991).

31. Zhang, S. \& Levy, P. M. Conductivity perpendicular to the plane of multilayered structures. \emph{J. App. Phys.} \textbf{69}, 4786 (1991).

32. Feldman, B. E., Krauss, B., Smet, J. H. \& Yacoby, A. Unconventional sequence of fractional quantum Hall states in suspended graphene. \emph{Science} \textbf{337}, 1196 (2012).

33. Bode, M. \emph{et al}. Magnetization-direction-dependent local electronic structure probed by scanning tunneling spectroscopy. \emph{Phys. Rev. Lett.} \textbf{89}, 237205 (2002).

34. Gould, C. \emph{et al}. Tunneling anisotropic magnetoresistance: a spin-valve-like tunnel magnetoresistance using a single magnetic layer. \emph{Phys. Rev. Lett.} \textbf{93}, 117203 (2004).

35. von Bergmann, K. \emph{et al}. Tunneling anisotropic magnetoresistance on the atomic scale. \emph{Phys. Rev. B} \textbf{86}, 134422 (2012).

36. Dzyaloshinskii, I. A thermodynamic theory of 'weak' ferromagnetism of antiferromagnets. \emph{J. Phys. Chem. Solids} \textbf{4}, 241 (1958).

37. Moriya, T. Anisotropic superexchange interaction and weak ferromagnetism. \emph{Phys. Rev.} \textbf{120}, 91 (1960).

38. Cr{\'e}pieux, A. \& Lacroix, C. Dzyaloshinskii-Moriya interactions induced by symmetry breaking at a surface. \emph{J. Magn. Magn. Mater.} \textbf{182}, 341 (1988).

39. Dup{\'e}, B., Hoffmann, M., Paillard, C. \& Heinze, S. Tailoring magnetic skyrmions in ultra-thin transition metal films. \emph{Nature Commun.} \textbf{5}, 4030 (2014).

40. Simon, E., Palot{\'a}s, K., R{\'o}zsa, L., Udvardi, L. \& Szunyogh, L. Formation of magnetic skyrmions with tunable properties in PdFe bilayer deposited on Ir(111). \emph{Phys. Rev. B} \textbf{90}, 094410 (2014).

41. Bauer, D., Mavropoulos, P., Zeller, R. \& Bl{\"u}gel, S. Non-collinear magnetism of Fe nano-islands on Ir(111): density-functional relativistic Green function calculations. \emph{Phys. Rev. B} Submitted (2015).

42. Bauer, D. \emph{Development of a relativistic full-potential first-principles multiple scattering Green function method applied to complex magnetic textures of nano structures at surfaces}, Ph.D. thesis, RWTH Aachen University (2014).

43. Tersoff, J. \& Hamann, D. R. Theory and application for the Scanning Tunneling Microscope. \emph{Phys. Rev. Lett.} \textbf{50}, 1998 (1983).

44. Vosko, S. H., Wilk, L. \& Nusair, M. Accurate spin-dependent electron liquid correlation energies for local spin density calculations: a critical analysis. \emph{Can. J. Phys.} \textbf{58}, 1200 (1980).

\section*{Acknowledgements}
We acknowledge the experimental group of Roland Wiesendanger at the University of Hamburg, whose ongoing experimental studies motivated in part this theoretical work. We thank the theoretical group of Stefan Heinze at the University of Kiel for insightful discussion. We recognize Sanjay K. Banerjee and L. Frank Register from the University of Texas at Austin for contributing expertise regarding realistic device designs related to advanced memory technologies. D.M.C. is supported by an NSF graduate fellowship. The authors also recognize funding sources from the DAAD Rise-\emph{Professional} Program and the HGF-YIG Program VH-NG-717 (Functional Nanoscale Structure and Probe Simulation Laboratory). 
\section*{Author Contributions}
D.M.C., M.B., and B.S. executed the \emph{ab initio} calculations. J.B. executed the model calculations. D.M.C. prepared the figures. S.L. conceived the research design and aims. D.M.C. prepared the manuscript. D.M.C., S.L., and S.B. analyzed the results and commented on the final manuscript.
\section*{Competing Financial Interests}
The authors declare no competing financial interests.
\clearpage

\renewcommand{\figurename}{\textbf{Figure S}}
\captionsetup[figure]{labelfont=bf,justification=raggedright}
\captionsetup[figure]{labelformat=simple, labelsep=none}
\section*{Supplementary Information}
\subsection*{S1 \---- Charge density and angular momentum of FM thin-film heterostructures}
Below we show tables containing information regarding the charge density, charge transfer, and spin and orbital magnetic moments of the near-surface atomic layers in the ferromagnetic (FM) state for each thin-film metallic system we studied, Pd/Fe/Ir(111) and Pd/Pd/Fe/Ir(111). The atomic number is $Z$, the total electronic charge is $n$, the charge transfer is $\Delta n$ = \big($n - Z$\big), the spin-moment is $\mathrm{M_S}$, and the orbital-moment is $\mathrm{M_L}$. $Z$, $n$, and $\Delta n$ are given in units of electrons while $\mathrm{M_S}$ and $\mathrm{M_L}$ are given in units of $\mu_\mathrm{B}$. We give the layers from top-down in the single-Pd system.  
\begin{center}
	\begin{tabular}{ l c c c c c c } 
	\hline\hline
	\multicolumn{7}{c}{\bfseries Pd/Fe/Ir(111)}	\\
	\hline\hline
	&\;		&	\quad$Z$\quad	&	\quad$n$\quad		&	\quad$\Delta n$\quad 	&	\quad$\mathrm{M_S}$\quad		&	\quad$\mathrm{M_L}$\quad			\\
	\hline
	&Pd		&	\quad46\quad	&	\quad45.72\quad	&	\quad$-$0.28\quad		&	\quad0.31\quad				&	\quad0.02\quad				\\
	&Fe		&	\quad26\quad	&	\quad26.18\quad	&	\quad0.18\quad			&	\quad2.70\quad				&	\quad0.10\quad					\\
	&Ir		&	\quad77\quad	&	\quad76.84\quad	&	\quad$-$0.16\quad		&	\quad0.02\quad				&	\quad$-$0.01\quad			        	\\
	\end{tabular}
\end{center}
We see that the spin magnetic moment of Fe is rather large (2.70 $\mu_\mathrm{B}$) and the orbital moment is not negligible (0.10 $\mu_\mathrm{B}$). Fe induces a sizable spin-moment in Pd of 0.31 $\mu_\mathrm{B}$ which \emph{stiffens} the isotropic exchange constant for first-nearest neighbors $J$ compared to the system$^{\mathrm{S}1,\mathrm{S}2}$ having no Pd-overlayer  ($J^{\mathrm{Pd/Fe/Ir}}$ = 14.7 meV vs. $J^{\mathrm{Fe/Ir}}$ = 5.7 meV). Pd is known to be a good Stoner system, i.e.\ it can easily develop a spin magnetic moment. Next we give a similar table for the double-Pd case (Pd/Pd/Fe/Ir):
\begin{center}
	\begin{tabular}{ l c c c c c c } 
	\hline\hline
	\multicolumn{7}{ c }{\bfseries Pd/Pd/Fe/Ir(111)}	\\
	\hline\hline
	&\;		&	\quad$Z$\quad	&	\quad$n$\quad		&	\quad$\Delta n$\quad 	&	\quad$\mathrm{M_S}$\quad		&	\quad$\mathrm{M_L}$\quad		\\
	\hline
	&Pd		&	\quad46\quad	&	\quad45.77\quad	&	\quad$-$0.23\quad		&	\quad0.08\quad				&	\quad0.01\quad					\\
	&Pd		&	\quad46\quad	&	\quad45.94\quad	&	\quad$-$0.06\quad		&	\quad0.28\quad				&	\quad0.02\quad				\\
	&Fe		&	\quad26\quad	&	\quad26.20\quad	&	\quad0.20\quad			&	\quad2.63\quad				&	\quad0.09\quad				\\
	&Ir		&	\quad77\quad	&	\quad76.83\quad	&	\quad$-$0.17\quad		&	\quad0.02\quad				&	\quad$-$0.01\quad					\\
	\end{tabular}
\end{center}
Interestingly in this case, the spin magnetic moments of Fe and the nearest-neighbor Pd did not change greatly compared to the values obtained for Pd/Fe/Ir. However, the charge transfer of the nearest-neighbor Pd-layer is now much smaller. Since the surface Pd-layer interacts with Fe indirectly through the inner Pd-layer, the induced spin magnetic moment is rather small ($0.08 \;\mu_\mathrm{B}$).
\subsection*{S2 \---- LDOS in single-Pd 2.2~nm skyrmions}
Let us now turn to the case of a single-Pd system but for a larger skyrmion, $D_{\mathrm{Sk}} \approx$ 2.2~nm, compared to the one shown in the main manuscript ($D_{\mathrm{Sk}} \approx 1.7$~nm, Fig.~3). In Fig.~S1, we plot the LDOS in a realistic nano-skyrmion in Pd/Fe/Ir(111) for the magnetically-active Fe-layer (\textbf{a}) and its corresponding vacuum sites (\textbf{b}), but with a slightly larger energy window $\left[-1.5, +1.5\right]$ eV as compared to the main text, where a window of $\left[-1.0, +1.0\right]$ eV was analyzed.

\setcounter{figure}{0}
\begin{figure*}[t!]
       \includegraphics[width=0.95\textwidth]{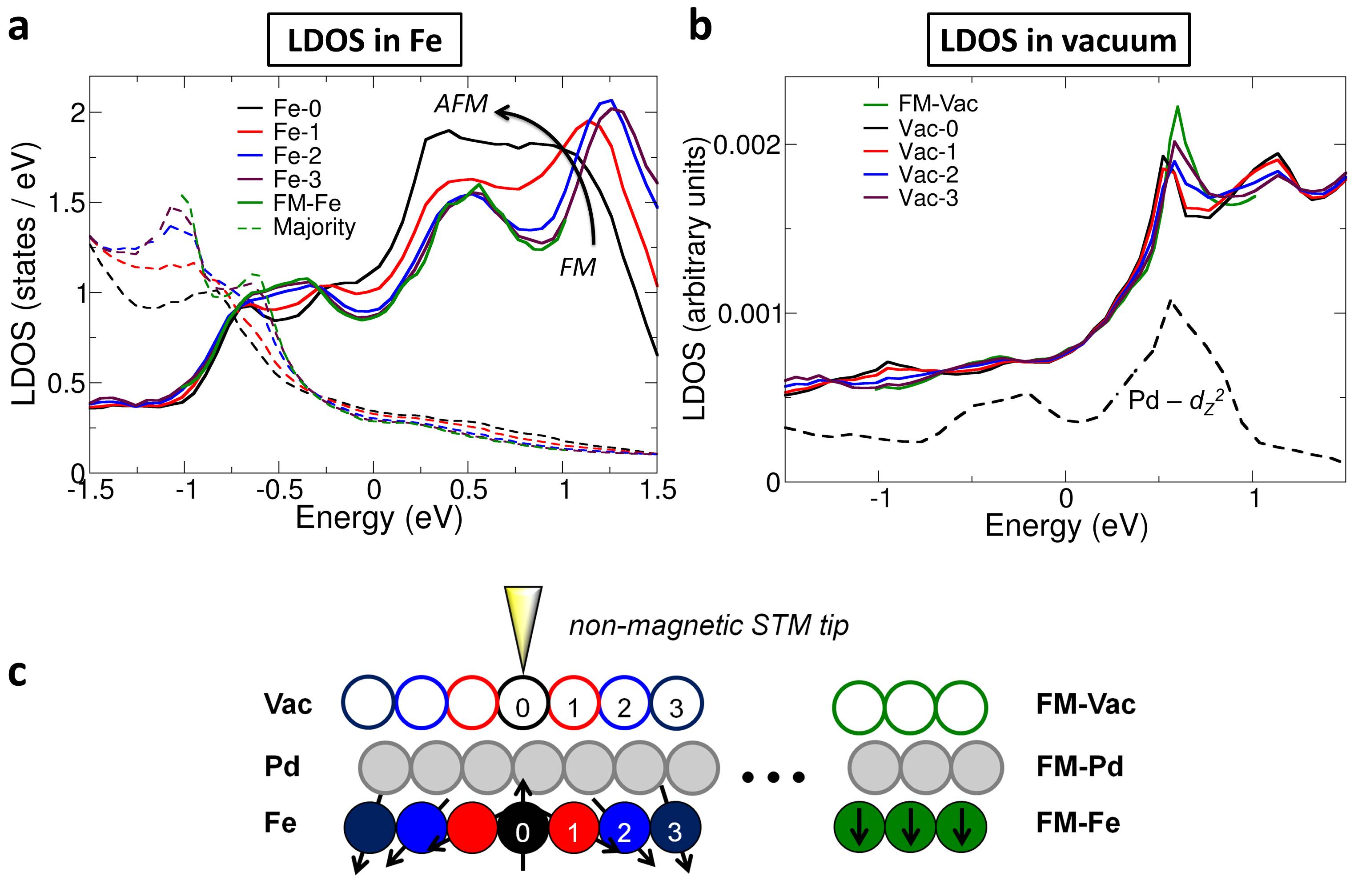}
	\caption{\label{fig:2_2ldos_pd} \textbf{\textbar\; LDOS in a 2.2~nm skyrmion for Pd/Fe/Ir(111). a}, With diameter $D_{\mathrm{Sk}} \approx$ 2.2~nm, shown is the LDOS in the Fe-layer resolved into minority (solid lines) and majority (dashed) spin-channels. \textbf{b}, LDOS in vacuum using arbitrary units so as to include the nature of the surface-layer Pd-$d_{z^2}$ state (dashed-black curve). \textbf{c}, The illustrative legend for (\textbf{a}, \textbf{b}) where the atoms are numbered and color-coded to identify the appropriate curves to the corresponding atoms in the skyrmion. The vacuum domains are represented by empty spheres. }
\end{figure*}

Similar to the case of the smaller skyrmion in the single-Pd system as shown in the main manuscript (Fig.~3a, FM-Fe, green curve), a large resonance peak is observed in the background-FM Fe-film at $0.5$~eV (Fig.~S1a, FM-Fe, green curve). But by increasing the energy window beyond $1.0$~eV, we clearly see a \emph{second} resonance appearing at $1.25$~eV in the nearly-FM atom Fe-3 at the edge of the skyrmion (Fig.~S1a, Fe-3, navy curve). The appearance of these two resonance peaks in tandem suggests a strong interaction due to ferromagnetic coupling. In other words, the resonant-peaks at $0.5$~eV and $1.25$~eV are actually the result of FM-interactions which have split a single resonance in two. An analogy can be made between these FM-splitted resonance-peaks and localized degenerate atomic orbitals at different sites which upon interaction split in energy between bonding- and antibonding-states. The essential physics of this process can be reproduced within an Alexander-Anderson model for $d$-$d$ hybridization between FM-Fe $d$-states$^{\mathrm{S}3,\mathrm{S}4}$ (see Supplementary Information S5).

The splitting in energy between the two resonance-peaks reduces when the rotation angle of the magnetic moment increases (Fig.~S1a, Fe-2, blue curve, and Fe-1, red curve). Finally, in the central spin-flipped atom of the skyrmion (Fig.~S1a, Fe-0, black curve), the splitting almost disappears as the two resonance-peaks quasi-merge. This is evidence of antiferromagnetic coupling between the central atom and its nearest-neighbors$^{\mathrm{S}3,\mathrm{S}4}$. We also reproduce this coupling within our Alexander-Anderson model in the antiferromagnetic case (see Supplementary Information S5). 

An important difference with respect to the center of the skyrmion (Fig.~S1a, Fe-0, black curve) is observed when comparing to the smaller $D_{\mathrm{Sk}} \approx$ 1.7~nm case (Fig.~3a, Fe-0, black curve).  In the smaller skyrmion, the quasi-merged resonance still shows two separate peaks, while in the larger skyrmion there is only a single, broad resonance peak. Thus we expect a difference in the TXMR signal in the larger skrymion compared to the $D_{\mathrm{Sk}} \approx$ 1.7~nm case. In general, there will be a weak size dependence of the spin-mixing signal, which should slowly decay as the diameter of the skyrmion is increased. In addition, from Fig.~S1b we can see that the vacuum resonance near 0.5~eV in the smaller defect (Fig.~3b, solid curves) survives in the larger skyrmion (Fig.~S1b, solid curves), and the TXMR remains detectible, as seen in Fig.~S6a.

To complete our analysis of the LDOS shown in Fig.~S1, we mention that with regards to magnifying the strength of the TXMR effect, we need not restrict ourselves only to the energy window near $0.5$~eV, but also look for other energy windows where a large change in the LDOS is observed as a function of position. We see this in the majority-states near $-1.0$~eV (Fig.~S1a, dashed-curves), where the spin-mixing effect is perhaps even stronger than the $0.5$~eV window composed of minority-states. Indeed, the TXMR signal is shown to be large for negative bias energies near $-1.0$~eV in this system (see Figs.~4a,c and S6a).

\subsection*{S3 \---- LDOS of skyrmions in the double-Pd systems}
In Fig.~S2, we plot the LDOS in a confined nano-skyrmion in Pd/Pd/Fe/Ir(111) with diameter $D_{\mathrm{Sk}} \approx$ 1.7~nm for the magnetically-active Fe-layer (\textbf{a}) and its corresponding vacuum sites (\textbf{b}).

\begin{figure*}[t!]
       \includegraphics[width=0.95\textwidth]{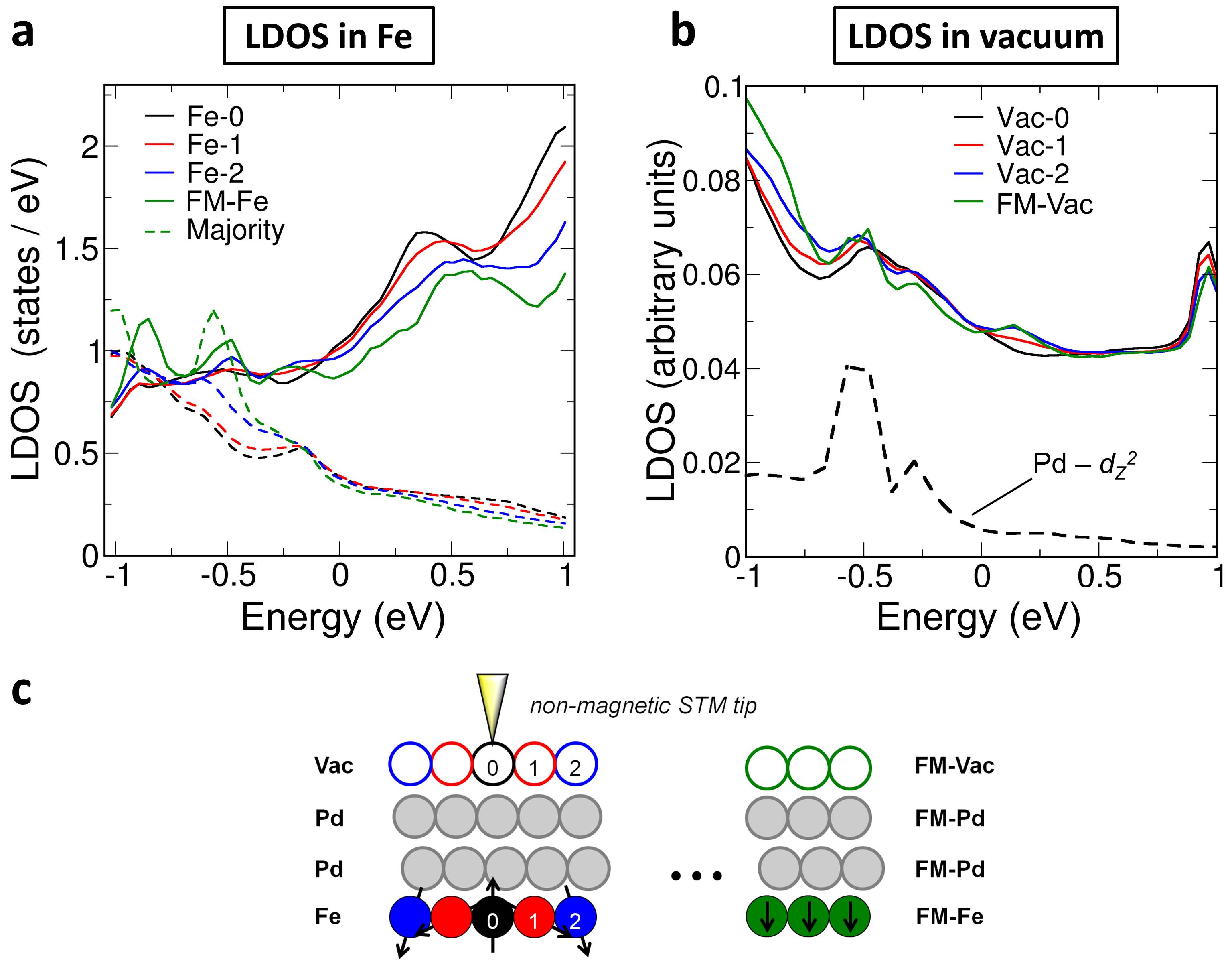}
	\caption{\label{fig:ldos_pd_pd} \textbf{\textbar\; LDOS in a realistic 1.7 nm skyrmion in Pd/Pd/Fe/Ir(111). a}, With diameter $D_{\mathrm{Sk}} \approx$ 1.7 nm, shown is the LDOS in the Fe-layer resolved into minority (solid lines) and majority (dashed) spin-channels. \textbf{b}, LDOS in vacuum using arbitrary units so as to include the nature of the surface-layer Pd-$d_{z^2}$ state (dashed-black curve). \textbf{c}, The illustrative legend for (\textbf{a}, \textbf{b}) where the atoms are numbered and color-coded to identify the appropriate curves to the corresponding atoms in the skyrmion. The vacuum domains are represented by empty spheres. }
\end{figure*}

Similar to the case of the smaller skyrmion in the single-Pd system as shown in the main manuscript (Fig.~3a, FM-Fe, green curve), a strong resonance-peak appears near $0.5$~eV composed of minority-spin channel $d$-states (Fig.~S2a, FM-Fe, green curve). This resonance-peak shifts in energy upon increased rotation of the magnetic moments moving towards the center of the skyrmion (Fig.~S2a, Fe-2, blue curve, Fe-1, red curve, and finally Fe-0, black curve). This observed change in the electronic structure as a function of position due to the spin-mixing of majority- and minority-states suggests there will be a TXMR signal in this system as shown previously for the single-Pd system. 

Contrary to the single-Pd case, however, the resonance in the hybrid Fe-Pd-$spd_{z^2}$ state (Pd-$d_{{z}^2}$ for short) observed in the Pd-overlayer around 0.5~eV vanishes, and instead appears near $-0.5$~eV (Fig.~S2b, dashed-black curve). The steepness of this resonance leads to a disturbance in the vacuum-LDOS as a function of position in slightly lower energies, near $-0.8$~eV (Fig.~S2b, solid curves). The character of the hybrid Pd-$d_{{z}^2}$ surface-state leads to a flat region in the vacuum-LDOS at positive energies. We then expect here a smaller TXMR signal in comparison to the single Pd-system for positive bias voltages. Therefore, in the double-Pd system, we suggest probing negative bias energies near $-0.8$~eV, where the spin-mixing effect is more significant, as exemplified in Fig.~4b,d.

\begin{figure*}[t!]
       \includegraphics[width=0.95\textwidth]{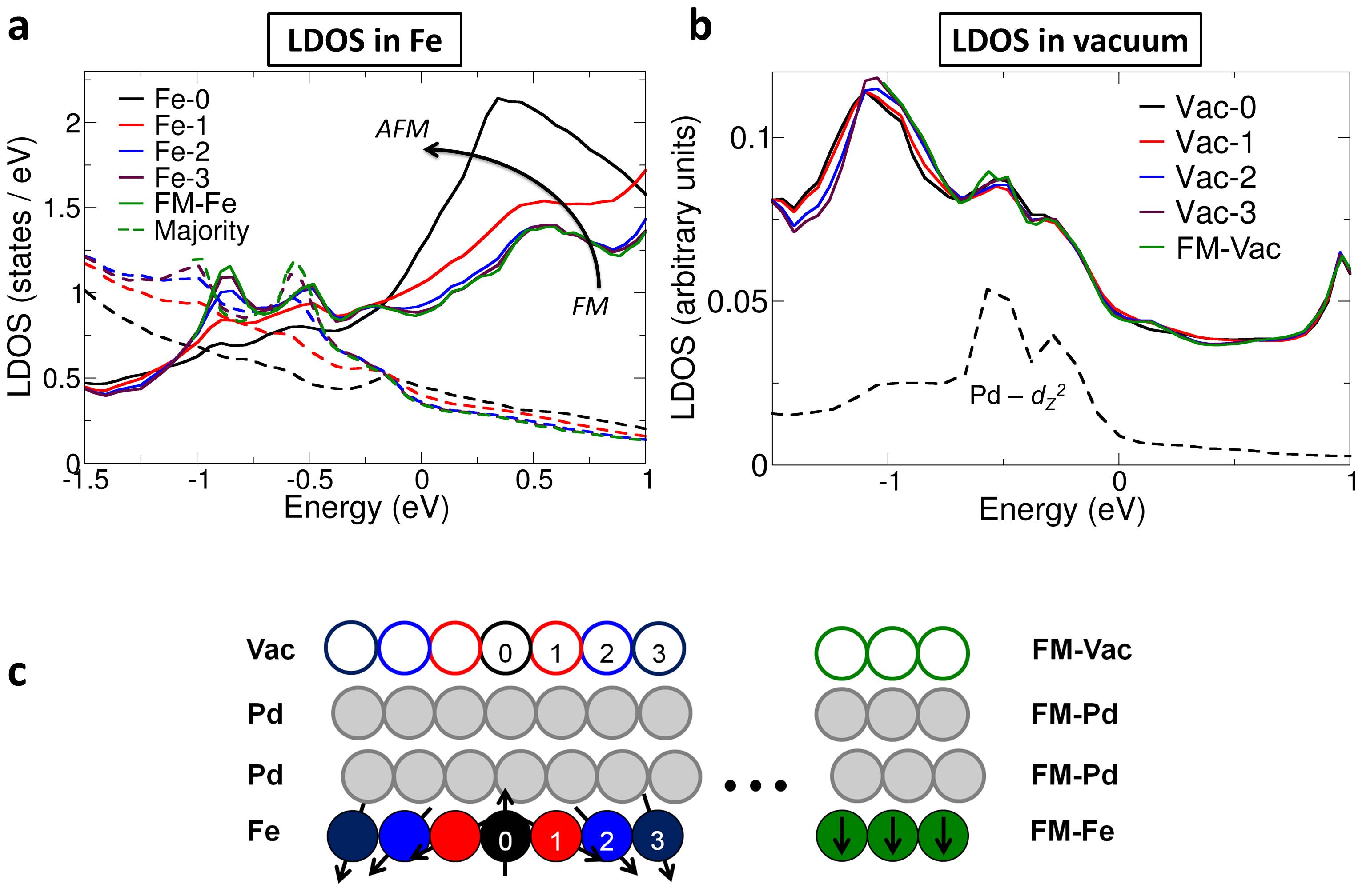}
	\caption{\label{fig:2_2_ldos_pd_pd} \textbf{\textbar\; LDOS in a 2.2 nm skyrmion in Pd/Pd/Fe/Ir(111). a}, With diameter $D_{\mathrm{Sk}} \approx$ 2.2 nm, shown is  the LDOS in the Fe-layer resolved into minority (solid lines) and majority (dashed) spin-channels. \textbf{b}, LDOS in vacuum using arbitrary units so as to include the nature of the surface-layer Pd-$d_{z^2}$ state. \textbf{c}, The illustrative legend for (\textbf{a}, \textbf{b}) where the atoms are numbered and color-coded to identify the appropriate curves to the corresponding atoms in the skyrmion. The vacuum domains are represented by empty spheres. }
\end{figure*}

In Fig. S\ref{fig:2_2_ldos_pd_pd}, we plot the LDOS for a larger skyrmion, $D_{\mathrm{Sk}} \approx$ 2.2~nm, in Pd/Pd/Fe/Ir(111). We use a similar labeling convention when decomposing the LDOS. Interestingly, the $d$-resonances are broader (Fig.~S3a, solid curves) than for the smaller skyrmion (Fig.~S2a, solid curves). In the center of the skyrmion, only one broad $d$-resonance is observed (Fig.~S3a, Fe-0, black curve). To reiterate, this indicates that the TXMR signal will be different upon increasing the diameter of any skyrmion since its electronic structure is modified upon increasing $D_{\mathrm{Sk}}$. From the shape of the vacuum-LDOS (Fig.~S3b, solid curves), the TXMR is nicely detectable at an energy range around $-0.8$~eV, as shown in Fig.~S6b, and in the same energy window as the smaller $D_{\mathrm{Sk}} \approx$ 1.7~nm Pd/Pd/Fe/Ir(111) case.

From Supplementary Sections S2 and S3 we can conclude that for each skyrmion in a Pd/Fe/Ir(111) or Pd/Pd/Fe/Ir(111) system, we can select an energy region in which a large tunneling spin-mixing magnetoresistance (TXMR) will be found. Due to the relevant energy of detection, one can distinguish skyrmions of different sizes, and discriminate between skyrmions in Pd-films of different thicknesses. Finally, and perhaps most importantly, the spatial-variation of the spin-mixing signal means that the \emph{internal} atomic structure of each individual skyrmion can be resolved. This means the TXMR can be used to visualize the size, shape, and structure of individaul defects. In devices based on spin-mixing, this also means that skyrmions can be used as nano-scopic information carriers, where the TXMR would be used to read the magnetic state-of-bit.

\subsection*{S4 \---- Angular dependence of the local-density-of-states in nano-skyrmions}
We now analytically derive the change in the LDOS at site $r$, denoted $\Delta$LDOS$(r)$, inside an axially symmetric skyrmion measured from the origin at $r=0$ as a function of energy and magnetization rotation direction defined by the unit vector of the 
magnetic moment $\mathbf{\hat{s}}(r)=\big(\sin\theta(r)\cos\phi(r), \sin\theta(r)\sin\phi(r), \cos\theta(r)\big)$. Examples of rotation parameters $\theta$ and d$\theta$ for given skyrmion magnetic moments can be found in Fig.~S5d.  When compared to the ferromagnetic state (with all moments pointing out-of-plane), there are two contributions to $\Delta$LDOS$(r)$, one due to spin-orbit interaction (SOI) and one due to non-collinearity (NC): $\Delta$LDOS$(r)$ = $\Delta$LDOS$^{SOI}(r)$ + $\Delta$LDOS$^{NC}(r)$. The contribution from SOI, the so-called anisotropic magnetoresistance, is well known$^{\mathrm{S}5-\mathrm{S}7}$: 
\begin{equation}
\Delta\mathrm{LDOS}^{SOI}(r, E, \theta) \propto A(r, E)\cdot\left[1 - \hat{s}^2_z(r)\right]  \;\;,
\end{equation}
where $A(r, E)$ is a coefficient depending on the site $r$ and energy $E$, and $\hat{s}_z(r)$ is the $z$-component of the spin-moment at site $r$. Thus upon including SOI, we expect, for example, a $\sin^2\theta(r)$ dependence, which contributes to $\Delta$LDOS$(r)$ in second-order.

The contribution from NC, intuitively, comes from the change in the electronic structure upon rotation of the magnetic moments at consecutive sites ($i$, $j$). For homogeneous magnetic spirals, a constant deviation in the LDOS from the ferromagnetic (FM) state will be observed for each atom in the spiral (ignoring SOI). In such a spiral, the smooth rotation of moments $\theta(i) \rightarrow \theta(j) = \theta(i)+\mathrm{d}\theta$ for each atom pair is a symmetry operation commuting with a Hamiltonian having translational invariance, making each atom equivalent, and the electronic structure the same for each atom in the spiral. However, upon transforming the spiral's pitch d$\theta\rightarrow\;$d$\theta^\prime$, one would find a \emph{different} constant deviation of the electronic structure from the FM-state, such that spirals of different pitch can be identified by their different magnitudes in $\Delta$LDOS$^{NC}$.

In skyrmions, however, the rotation of the magnetic moments is not homogeneous, i.e.\ d$\theta$ is not constant for all atom pairs. Thus there will be a site-dependent deviation in the LDOS among the atoms inside a skyrmion with respect to each other. In what follows, we demonstrate that these deviations are a complex function of the rotation angles, which will depend on the details of the electronic structure, the energy probed, and even the size of the skyrmions.

To derive $\Delta \mathrm{LDOS}^{NC}(r)$, we utilize multiple scattering theory: $G$ is the Green function describing the whole system upon rotation of the magnetic moments, and $g$ is the Green function describing the initial FM-state. The Green function will be used to evaluate the change in the LDOS induced by the rotation of the magnetic moments:
\begin{equation}
\Delta \mathrm{LDOS}^{NC}(r, E) = -\frac{1}{\pi} \ \Im \bigg\{ \mathrm{Tr_{LS}}\big[ \Delta G^{rr}(E) \big] \bigg\} \;\; ,
\label{LDOS}
\end{equation}
as given in a matrix notation where a trace over orbital and spin angular momenta has to be performed. $G$ can be evaluated via the Dyson equation connecting the non-collinear state to the ferromagnetic one:
\begin{equation}
G = g+ g\Delta V G = g + g \Delta V g + g \Delta V g \Delta V g + ....
\label{dyson}
\end{equation}
where $\Delta V$ describes the change of the potential upon rotation of the magnetic moments. It can be expressed as:
\begin{equation}
\Delta V(r) = V_{\mathrm{diff}}(r)  \left( \mbox{\boldmath $\sigma$} \cdot  \mathbf{\hat{s}}(r)  - {\sigma}_z \right) \;\; ,
\end{equation}
where $\mbox{\boldmath $\sigma$}$ is the vector of Pauli matrices, and $V_{\mathrm{diff}}$ is the difference of the two spin components of the ferromagnetic potential $(V^{\uparrow}_0-V^{\downarrow}_0)/2$.

We execute a similar expansion for the ferromagnetic initial Green function matrix $g$:
\begin{equation}
g = g_{\mathrm{sum}}  1_2 + g_{\mathrm{diff}}  {\sigma}_z \ \;\; ,
\end{equation}
where $1_2$ is the $2\times2$ identity matrix.

Let us evaluate the first-order and second-order terms contributing to the Dyson Equ.~\ref{dyson}:
\begin{equation}
\Delta G^{rr} =  \sum_i g^{ri} \Delta V^i g^{ir} + \sum_{ij} g^{ri} \Delta V^i g^{ij} \Delta V^j g^{jr} + ....
\label{Born}
\end{equation}
where $i$ and $j$ are sites surrounding site $r$, or can be the site $r$ itself. Since the trace over spin has to be performed, we will focus only on the terms that in the end will contribute to Equ. \ref{LDOS}. We use a pair of useful properties of the Pauli matrices:
\beq
\mathrm{Tr_S}\big[\sigma_x\big] = \mathrm{Tr_S}\big[\sigma_y\big] = \mathrm{Tr_S}\big[\sigma_z\big] =0 \;\; ,
\eeq
and
\begin{equation}
(\mbox{\boldmath $\sigma$} \cdot \mathbf{\hat{s}})(\mbox{\boldmath $\sigma$} \cdot \mathbf{\hat{s}}^\prime) = \mathbf{\hat{s}} \cdot \mathbf{\hat{s}}^\prime + \mathrm{i} \ \mbox{\boldmath $\sigma$} \ \cdot  (\mathbf{\hat{s}} \times \mathbf{\hat{s}^\prime}) \; \;,
\end{equation}
where $\mathrm{i}$ is the imaginary unit. After simplifying, we find the following result:
\beq
-\frac{1}{\pi}\Im \bigg\{ \mathrm{Tr}_\mathrm{{LS}}\big[g^{ri} \Delta V^i g^{ir}\big]\bigg\} &=& B^{rir}(E)\left[1-\hat{s}_z(i)\right] 
\eeq 
where the coefficient $B$ is given by 
\beq
 B^{rir} = \frac{2}{\pi} \Im \bigg\{ \mathrm{Tr}_\mathrm{{L}}\big[ g_{\mathrm{sum}}^{ri}    V_{\mathrm{diff}}^i  g_{\mathrm{diff}}^{ir} + g_{\mathrm{diff}}^{ri}   V_{\mathrm{diff}}^i   g_{\mathrm{sum}}^{ir}\big] \bigg\} \;\; .
\label{Born11}
\eeq
In other words, the first sum in Equ.~\ref{Born} leads to a behavior like $(1-\cos\theta_i)$.

The second-order term is given by
\begin{eqnarray}
-\frac{1}{\pi}\Im \bigg\{\mathrm{Tr}_{\mathrm{LS}}\big[ g^{ri} \Delta V^i g^{ij}\Delta V^j g^{jr}\big]\bigg\}  =  
 C^{rijr}\big[  \mathbf{\hat{s}}(i)\cdot  \mathbf{\hat{s}}(j)  - \big(\hat{s}_z(i) + \hat{s}_z(j)\big) 
 +1  \big] \;\;,
\label{Born2}
\end{eqnarray}
where the coefficient $C$ is related to the Green functions and $V_{\mathrm{diff}}$ by:
\begin{eqnarray}
C^{rijr} = \nonumber -\frac{2}{\pi}\Im \bigg\{ \mathrm{Tr}_{\mathrm{L}} \big[  g_{\mathrm{sum}}^{ri}  V_{\mathrm{diff}}^i g_{\mathrm{sum}}^{ij}V_{\mathrm{diff}}^jg_{\mathrm{sum}}^{jr} \;+\; g_{\mathrm{diff}}^{ri} V_{\mathrm{diff}}^i g_{\mathrm{diff}}^{ij}V_{\mathrm{diff}}^j g_{\mathrm{sum}}^{jr} 
	     \\+\;  g_{\mathrm{sum}}^{ri}V_{\mathrm{diff}}^i g_{\mathrm{diff}}^{ij}V_{\mathrm{diff}}^j g_{\mathrm{diff}}^{jr} \;+\;
           g_{\mathrm{diff}}^{ri}V_{\mathrm{diff}}^ig_{\mathrm{sum}}^{ij}V_{\mathrm{diff}}^jg_{\mathrm{diff}}^{jr}\big] \bigg\} \;\;.
\end{eqnarray}

Thus we obtain a dependence on the dot product of the unit vectors of the magnetic moments ($1-\cos\mathrm{d}\theta \cos\mathrm{d}\phi$) and a contribution depending only on the $z$-components of the unit vectors of the magnetic moments. 

We have thus demonstrated that due to NC, the dependence of the change in the LDOS with respect to the ferromagnetic state upon rotation of the magnetic moments is not trivial, and will have terms depending on the dot product between magnetic moments, contrary to the contribution coming from SOI. The non-collinear contribution is then
\begin{eqnarray}
\nonumber &\Delta\mathrm{LDOS}^{NC}(r, E, \{ \mathbf{s}\}) = \\ &\sum\limits_{i} B^{rir}(E)\big(1-\hat{s}_z(i)\big) +\;  \sum\limits_{ij} C^{rijr}(E) \big[\mathbf{\hat{s}}(i)\cdot  \mathbf{\hat{s}}(j)  - \big(\hat{s}_z(i) + \hat{s}_z(j)\big) 
 +1 \big] \;\;,
\label{final_model1}
\end{eqnarray}
where $\{\mathbf{s}\}$ is the spin configuration. Of course, depending on the details of the electronic structure and strength of perturbation related to the non-collinearity, higher-order terms can be important and have to be included in Equ.~\ref{final_model1}.

Combining $\Delta\mathrm{LDOS}^{NC}$ and $\Delta\mathrm{LDOS}^{SOI}$, in the next section we will fit the change in the LDOS in terms of trigonometrical functions that depend on the rotation angles of the magnetic moments. We will apply these fits to our \emph{ab initio} results as well as to an extended Alexander-Anderson model used to interpret the variation of the LDOS resonance-splitting upon rotation of the magnetic moments on neighboring sites.

\subsection*{S5 \---- Two-atom extended Alexander-Anderson model using Green functions}
We wish to estimate the change in the LDOS and qualitatively understand the shifting in energy of resonant $d$-states in Fe as a function of the rotation angle between adjacent moments. To this end, we consider for simplification two magnetic atoms ($i$, $j$) = (1, 2) each having one localized orbital 
$d_{z^2}$ whose single-particle eigenenergy is centered about $E=\epsilon$. The initial Hamiltonian describing this model is diagonal in spin-space. We could also consider an orbital of the type $d_{xz}$ in order to address the coupling induced by SOI between the $d_{z^2}$ and $d_{xz}$, as done by Caffrey \emph{et al}$^{\mathrm{S}8}$. However, since the impact of SOI on the LDOS has already been discussed by others, we focus here on the impact of NC on the LDOS. We study the $\Delta$LDOS as we vary d$\theta = \theta_1 - \theta_2$ between the two atoms. We restrict the hopping from atom-to-atom to non-spin-flip processes, characterized by the interaction parameter $V_\mathrm{hop}$. 

In terms of Green functions, the following equation gives the LDOS for site 1:
\begin{equation}
\mathrm{LDOS}(1; E, \{ \mathbf{s}\})= -\frac{1}{\pi} \Im\bigg\{ \mathrm{Tr_S}\big[G_{11}(E)\big]\bigg\} =  -\frac{1}{\pi} \Im\bigg\{ \mathrm{Tr_S}\left[{E-H +i\Gamma}\right]^{-1}\bigg\}_{11} \;\;,
\end{equation}
where $\Gamma$ takes care of the broadening of the states. Instead of solving exactly the previous equation, one could also use perturbation theory, as described in the previous section S4, simplifying Equ.~\ref{final_model1} to:
\beq
\Delta\mathrm{LDOS}^{NC}(1; E, \{ \mathbf{s}\}) =  D(E)\cdot\left(1-\cos\mathrm{d}\theta\right),
\label{final_model2}
\eeq
where $D = B^{121} + C^{1221}$.

\begin{figure*}[t!]
       \includegraphics[width=0.95\textwidth]{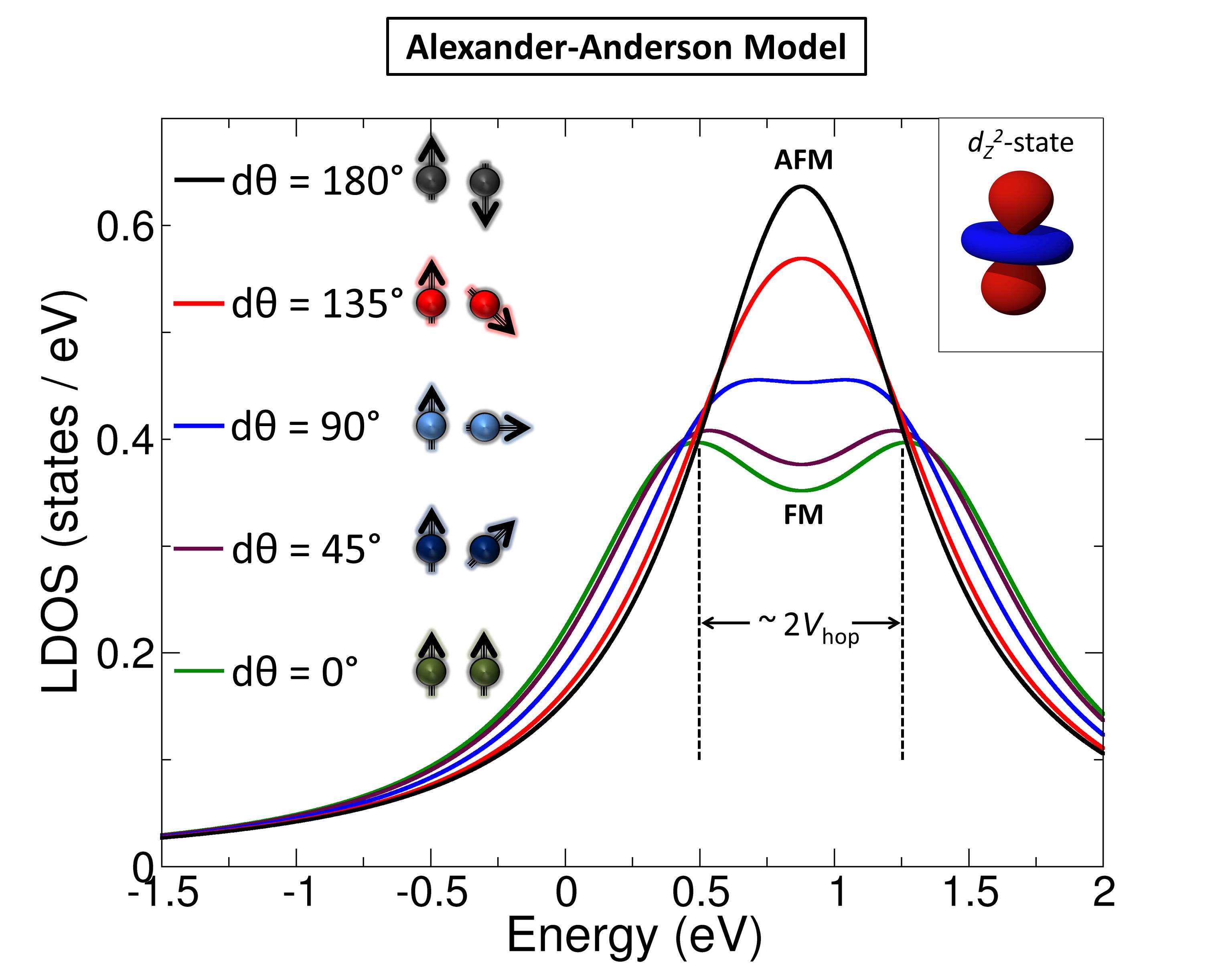}
	\caption{\label{fig:alex_model} \textbf{\textbar\; Alexander-Anderson prediction for $d$-$d$ hybridization. } Beginning from d$\theta$=0\textdegree, the ferromagnetic state imposes a 
hybridization between the two-orbitals localized at the same energy. This produces a splitting into a bonding- and antibonding-state (green curve). Upon rotation, the atoms eventually become antiferromagnetically coupled (black curve). The change in the LDOS can be qualitatively estimated with a $\cos$(d$\theta$) fitting parameter.}
\label{fig:modell}
\end{figure*}

The energy of the resonant $d$-states, their width, and splittings come from our first-principles calculations, e.g.\ Fig.~S1a (atom Fe-3, navy curve). To obtain the proper splitting we choose a hopping parameter $V_{\mathrm{hop}}\approx$ 300-400 meV. We show the resulting LDOS in Fig.~S\ref{fig:modell} for five different rotation angles d$\theta$. There we reproduce the splitting of the resonance-peaks that we have seen in our first-principles calculations, where $d$-$d$ hybridization is important, as seen in e.g.\ Fig.~3a of the main text, or Figs.~S1a, S2a, and S3a of the Supplementary Information. 

Interestingly, for the extreme cases of a ferromagnetic state and an antiferromagnetic state, we recover the expectations of the Alexander-Anderson model$^{\mathrm{S}3,\mathrm{S}4}$. Indeed, when the two atoms are in the ferromagnetic state (d$\theta=0$\textdegree), the $d$-$d$ hybridization leads to 
a splitting of the original single orbital into bonding- and antibonding-states, seen as broad resonances near $0.5$~eV and $1.25$~eV, respectively (Fig.~S4, green curve). The splitting is then given by $2 V_{\mathrm{hop}}$. 
If the magnetic state is antiferromagnetic (d$\theta=180$\textdegree), there is repulsion between the minority and majority spin-states leading to a shift given by 
$V_{\mathrm{hop}}/(E^{\downarrow}_{i,dz^2}-E^{\uparrow}_{i,dz^2})$. In our simple model, $E^{\downarrow}_{i,dz^2}-E^{\uparrow}_{i,dz^2}$ is extremely large, thus the shift is not observed in Fig.~S4. It is interesting to see how the splitting between the resonance-peaks decreases upon rotation of the magnetic moments until they merge to a single resonance in the antiferromagnetic case (Fig.~S4, black curve). This is in accordance with the behavior of the LDOS calculated from first-principles for different sized skyrmions in the two systems Pd/Fe/Ir (Figs.~3a and S1a) and Pd/Pd/Fe/Ir (Figs.~S2a and S3a).

Next, we wish to  estimate the change in the LDOS as a function of rotation as previously discussed, which leads to the TXMR signal. In Fig.~S5 we plot the change in LDOS compared against the background-FM for the model along with our \emph{ab initio} results in Pd/Pd/Fe/Ir for skyrmions 1.7~nm and 2.2~nm in diameter, respectively. In addition we show fitted functions against $\theta$ and d$\theta$. Interestingly, a good fit to the change in LDOS shown in Fig.~S5a,b 
is found by considering 
\begin{equation}
\Delta\mathrm{LDOS}(r; E, \{ \mathbf{s}\}) \approx A(E)\cdot \sin^2\theta(r) +  D(E)\cdot \big[1 -\cos\theta(r)\big]  \;\;
\label{eq:model}
\end{equation}
instead of strictly employing the terms given by Equ.~\ref{final_model1}. This result is similar to what we found in the simple two-orbital Alexander-Anderson model (Equ.~\ref{final_model2}). In such systems, the contribution from all spin-moment dot products (see Equ.~\ref{Born2}) behaves on \emph{average} like $\cos(\theta)$. This is naturally satisfied in the two-orbital model.

\begin{figure*}[t!]
       \includegraphics[width=0.95\textwidth]{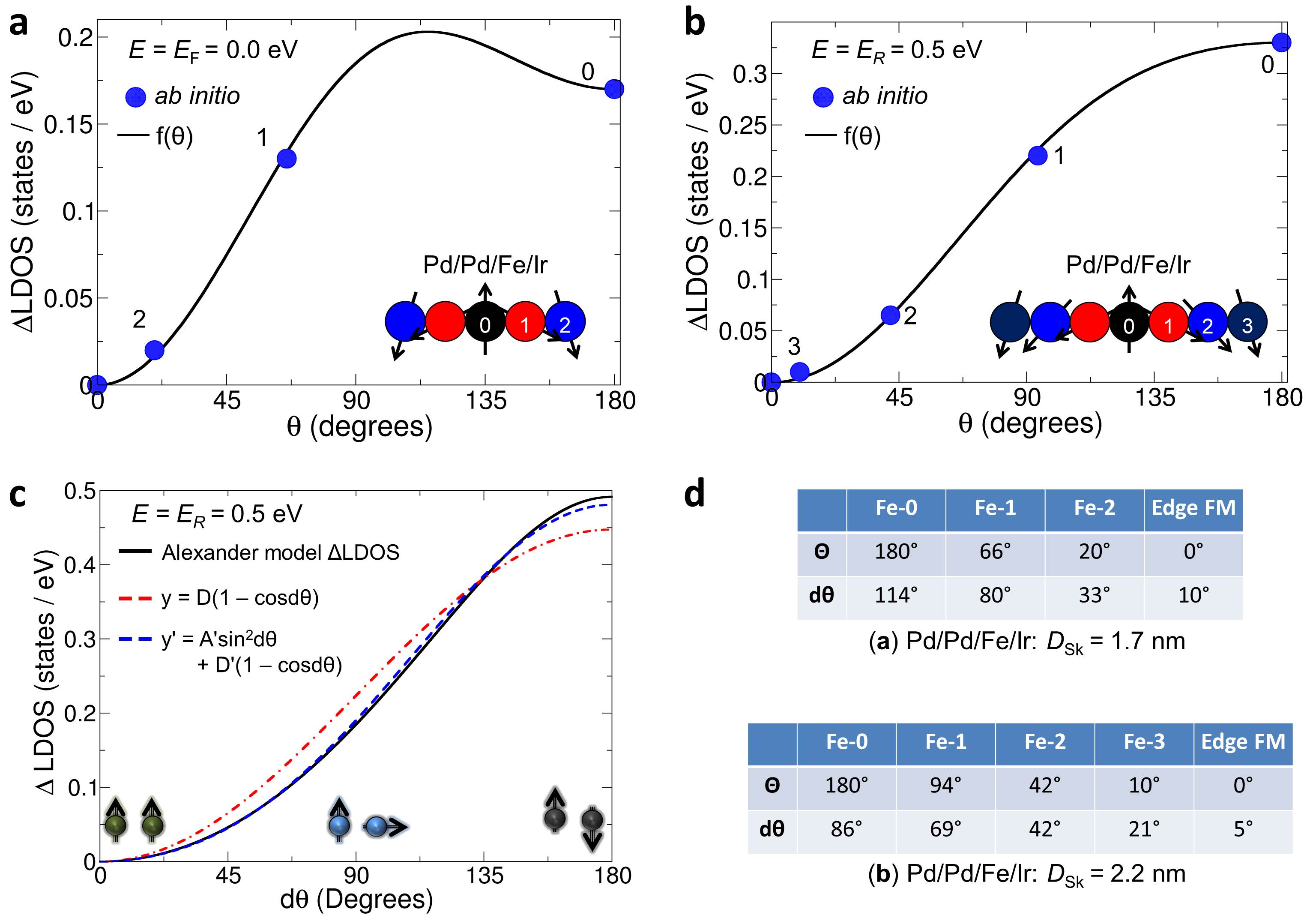}
	\caption{\label{fig:moder} \textbf{\textbar\; Energy-dependent change in LDOS in confined skyrmions. a,} Atom-by-atom \emph{ab initio} results (blue dots) of the total change in LDOS relative to the background-FM including SOI in 1.7 nm diameter skyrmion in Pd/Pd/Fe/Ir fitted to $f(\theta)$ (solid curve). The $f(\theta)$ in (\textbf{a,b}) refers to Equ.~16, while the numbers labeling each dot respresent the corresponding atom in the inset extending radially from the skyrmion core. \textbf{b,} Again, first-principles results (blue dots) of the atomistic change in LDOS but for a 2.2 nm skyrmion in Pd/Pd/Fe/Ir fitted against $f(\theta)$. \textbf{c,} Fitting the Alexander-Anderson model for the non-collinear component of the $\Delta$LDOS. \textbf{d,} Rotation parameters as found from the first-principles calculations, referring to (\textbf{a}) and (\textbf{b}). }
\end{figure*}

In Fig.~S5a the change in LDOS at the Fermi energy $E=E_{F}=0$ is depicted. The blue dots represent the total $\Delta$LDOS given from our \emph{ab initio} calculations for each atom extending radially from the skyrmion core. The fitting-function $f(\theta)$ is Equ.~\ref{eq:model} and shown to be effective in fitting the first-principles data. We note that the change in the LDOS at the Fermi energy was not large. This is different than the situations in Fig.~S5b,c where we probed the energy resonances near 0.5 eV, and a larger change in the LDOS was induced. In Fig.~S5b, the fitting function $f(\theta)$ is again Equ.~\ref{eq:model} and shown to be effective in fitting the \emph{ab initio} data. 

In Fig.~S5c, we plot the $\Delta$LDOS as given by the two-atom Alexander-Anderson model (black curve) with two fitting-functions. The first-order fit (red-dashed curve) based on Equ.~\ref{final_model2} is shown to be slightly inaccurate when fitting the model $\Delta$LDOS. Instead a higher-order term is needed to fit the data (blue-dashed curve). Thus we learned from the Alexander-Anderson model that depending on the probed energy, additional terms can be needed to improve the fit. A term proportional to $\sin^2(\theta)$, similar to the one expected when SOI is included, improves considerably the fit to the non-collinear contribution to $\Delta$LDOS. For example, near energy resonances, especially if they are sharp, the perturbative expansion up to second-order from Supplementary Section S4 begins to break down and higher-order terms are necessary to improve the fitting-functions. 
\subsection*{S6 \---- TXMR in larger 2.2~nm skyrmions}
Next we plot the TXMR signal spacially-resolved for the 2.2~nm skyrmions in both systems. The spin-mixing signals retain the same general features and shapes as in the smaller defects (Fig.~4c,d). The important peak near 0.5~eV in the Pd/Fe/Ir case remains (Fig.~S6a), along with another strong peak near $-1.0$~eV. In the case of Pd/Pd/Fe/Ir (Fig.~S6b), the TXMR has the strongest signal near $-0.8$~eV, as before. Therefore one could infer from Fig.~S6 that as the diameter of skyrmionic quasiparticles is increased, the spin-mixing effect not only survives, but also that specific locations of strong TXMR signals remain in similar energy windows as in the smaller structures. 
\begin{figure*}[t!]
       \includegraphics[width=0.95\textwidth]{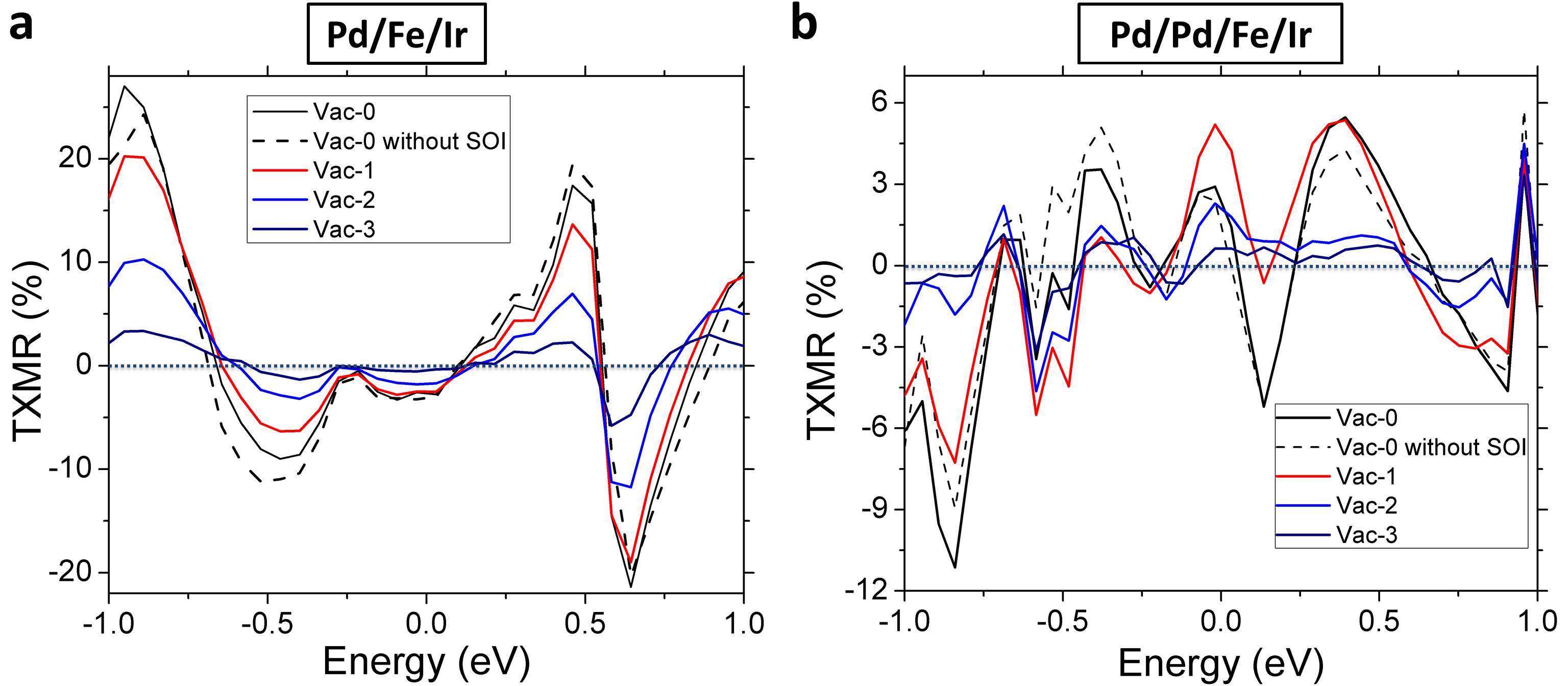}
	\caption{\label{fig:2_2txmr} \textbf{\textbar\; Spacially-varying TXMR in 2.2 nm skyrmions. a}, With diameter $D_{Sk} \approx$ 2.2 nm, we show the TXMR in vacuum for Pd/Fe/Ir. The vacuum resonance near 0.5 eV (Fig.~S1b) contributes a strong peak in the TXMR near the same energy. \textbf{b}, Again but in Pd/Pd/Fe/Ir, where the peak-signal strength appears near $-0.8$~eV, as predicted by Fig.~S3b. }
\end{figure*}

\subsection*{S7 \---- Skyrmion racetracks for dense magnetic memories}
Spin-transfer torque magnetic random access memory (STT-MRAM) circuits reliably read-out bit-states depending on a tunneling magnetoresistance anisotropy of $\sim$30-50\% in some structures$^{\mathrm{S}9}$, with a hope to achieve a magnetoresistance ratio $R_{\mathrm{ON}}/R_{\mathrm{OFF}}\approx200\%$ by 2022$^{\mathrm{S}10}$.  A TXMR effect as large as $\sim$20$\%$ as we have shown in this work should be enough to provide adequate read-margin for scaled technologies, and is larger than the $<2\%$ change in resistance found in widespread commercially-used hard disk read heads based on anisotropic magnetoresistance alone$^{\mathrm{S}11}$. Smaller changes in magnetoresistance just means there should be a more sensitive read-out circuit. Typically this means a few extra control- and boost-transistors and does not substantially increase the footprint of the memory, i.e. incorporating more sensitive read hardware does not degrade packing density considerably.

In potential skyrmion-based devices using CPP-TXMR, a $R_{\mathrm{ON}}/R_{\mathrm{OFF}}\approx120\%$ could feasably be well-worth the tradeoff when considering the possible performance gains with regards to: (1) potentially very low power dissipation due to small currents needed to manipulate the magnetic textures; (2) fast speed operation due to reduced read/write latencies associated with nano-scopic size; and (3) large increases in packing density. Let's consider points (2) and (3) in greater detail.

One issue with racetrack memories is that they are not \emph{random access}. In a random access memory (RAM), any read/write operation can access any bit with roughly the same access time since the word and bit access lines (WL and BL), which are connected to the set/reset elements and read elements, are also connected in parallel with the individual memory cells (see Fig.~S7a). In a racetrack memory, the situation is different. In practice, if a read/write were requested for an address whose representative bit were at the end of the racetrack, there would be additional latency associated with moving subsequent magnetic domains out the way until the requested bit were under the read or write device (see Fig.~S7b). But does this make racetrack memory intrinsically slower than RAM? 

\begin{figure*}[t!]
       \includegraphics[width=0.95\textwidth]{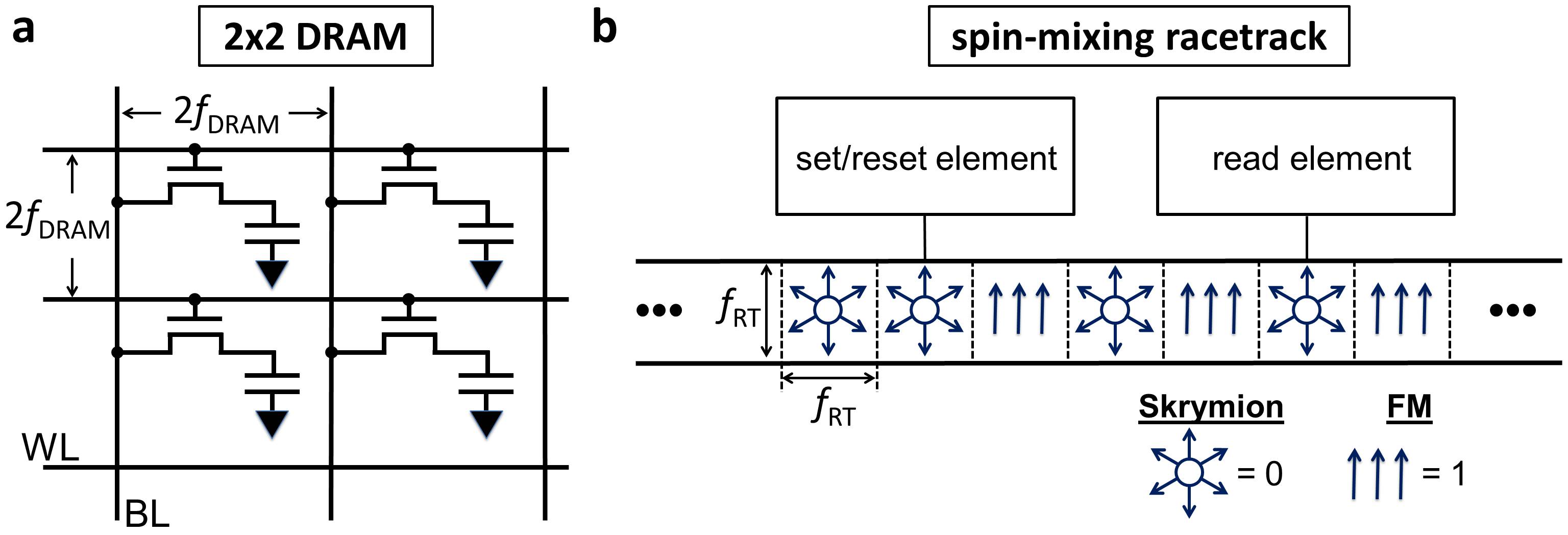}
	\caption{\label{fig:rt} \textbf{\textbar\; Concept spin-mixing magneto-memory versus DRAM. a}, 2$\times$2 1-T 1-C DRAM unit. In a random access memory, any memory cell can be accessed in roughly the same time since the WLs and BLs are connected to each bit in parallel. The ideal minimum packing footprint goes as $4f_{\mathrm{DRAM}}^2/\mathrm{bit}$ in units of area, where $f_{\mathrm{DRAM}}$ is the node generation. \textbf{b}, Spin-mixing racetracks are not random access but acquire a latency associated with moving the bits sequentially out of the way to reach the requested address. However, with realistic skyrmion velocities, the access time could be roughly the same as in DRAM, while dramatically increasing the packing density.}
\end{figure*}

This is not clear. It depends on the velocities of the skyrmions in the racetrack. Consider 200~nm length racetracks populated with 2 nm in diameter quasiparticles in an array of 100 domains (or 100 bits). With velocity$^{\mathrm{S}12}$ $v_\mathrm{Sk}$ = 100 m$/$s , the furthest bit would acquire an additional delay in time $t_\mathrm{delay}$
\be 
t_\mathrm{delay} = \frac{L_{\mathrm{RT}}}{v_\mathrm{Sk}} \approx \frac{\left(2\frac{\mathrm{nm}}{\mathrm{Sk}}\right)\left(1\frac{\mathrm{Sk}}{\mathrm{bit}}\right)\left(100\;\mathrm{bits}\right)}{100\left(\frac{\mathrm{m}}{\mathrm{s}}\right)} = 2 \;\mathrm{ns} \;\; ,
\ee
a small number compared to the total access time needed to complete the read or write operation, which in modern dynamic RAM (DRAM) is in the range 20\---50 ns$^{\mathrm{S}13}$. We do note, however, that larger in-plane currents will be required to accelerate the quasiparticles up to a velocity such as $v_\mathrm{Sk}$ = 100 m$/$s, meaning there will be a tradeoff between $t_\mathrm{delay}$ and power consumption. 

Regardless, by incorporating skrymion racetracks based on spin-mixing, there seems to be at first glance negligible additional acquired access latency \--- in fact, we may learn to find in the end that racetracks can be potentially \emph{faster} than traditional RAM in certain geometries and biasing conditions, due to the nano-scopic size of the skyrmion quasiparticles.

With regards to circuit layouts, it is clear that moving to racetracks will provide large gains in packing density. As an example, let's compare the workhouse 1-transistor 1-capacitor (1-T 1-C) DRAM unit to our racetrack-based spin-mixing magneto-memory. Considering the lithographical node generation, or minimum feature size, $f$, DRAM memory minimum packing requirements for a single bit's footprint in the \emph{ideal} case goes as surface area SA$_{\mathrm{DRAM}} \approx 4f^2$ per bit (Fig.~S7a). The current technology node for DRAM in 2015 is given as $f_{\mathrm{DRAM}} \approx 22$~nm by the International Technology Roadmap for Semiconductors (ITRS) $^{\mathrm{S}10}$. In a skyrmion racetrack, the potential effective per-bit SA could be reduced to possibly SA$_\mathrm{RT}\approx f^2_{\mathrm{RT}}$ per bit, where $f_{\mathrm{RT}}$ is the diameter of magnetic quasiparticles in the racetrack (Fig.~S7b). In our study thus far we considered realistic skyrmions of order $D_{\mathrm{Sk}}\approx$ 2~nm. Comparing against current DRAM arrays, we define the potential gain in packing density $\Gamma$ as
\be
\Gamma = \frac{SA_{\mathrm{DRAM}}}{SA_{\mathrm{RT}}} \approx \frac{4f_{\mathrm{DRAM}}^2}{f_{\mathrm{RT}}^2} = \frac{(4)(22\;\mathrm{nm})^2}{(2\;\mathrm{nm})^2} \approx 500 \;\;.
\ee
Even including more intricate sense amplifiers for read/write operations in skyrmion racetracks, a conservative estimate gives the increase in packing density around 10$\times$, with an upper limit around 500$\times$ if incorporating ultimately-scaled sub-5~nm nanoskyrmions and common fabrication techniques.  

Possibly the packing density could even be larger introducing \emph{vertical} racetracks$^{\mathrm{S}14}$, which are difficult to fabricate thus far.
\subsection*{Supplementary References}
S1. Dup{\'e}, B., Hoffmann, M.,  Paillard, C. \& Heinze, S. Tailoring magnetic skyrmions in ultra-thin transition metal films. \emph{Nature Commun}. \textbf{5}, 4030 (2014).

S2. Heinze, S. \emph{et al.} Spontaneous atomic-scale magnetic skyrmion lattice in two dimensions. \emph{Nature Phys.} \textbf{7}, 713 (2011).

S3. Oswald, A., Zeller, R.,  Braspenning, P.J. \& Dederichs, P. H. Interaction of magnetic impurities in Cu and Ag. \emph{J. Phys. F. Met. Phys.} \textbf{15}, 193 (1985).

S4. Alexander, S. \& Anderson, P. W. Interaction Between Localized States in Metals. \emph{Phys. Rev. A} \textbf{133}, 1594 (1964).

S5. Bode, M. \emph {et al.} Magnetization-direction-dependent local electronic structure probed by scanning tunneling spectroscopy. \emph{Phys. Rev. Lett.} \textbf{89}, 237205 (2002).

S6. Gould, C. \emph {et al.} Tunneling anisotropic magnetoresistance: a spin-valve-like tunnel magnetoresistance using a single magnetic layer. \emph{Phys. Rev. Lett.} \textbf{93}, 117203 (2004).

S7. von Bergmann, K. \emph{et al.} Tunneling anisotropic magnetoresistance on the atomic scale. \emph{Phys. Rev. B} \textbf{86}, 134422 (2012).

S8. Caffrey, N. M., \emph{et al.} Tunneling anisotropic magnetoresistance effect of single adatoms on a noncollinear magnetic surface. \emph{J. Phys.: Condens. Matter} \textbf{26}, 394010 (2014).

S9. Tehrani, S. T. \emph{et al.} Magnetoresistive random access memory using magnetic tunnel junctions. \emph{ Proceedings of the IEEE}
\textbf{91}, 703 (2003).

S10. See www.itrs.net for the latest 2013 edition of the technology roadmap for current and future logic and memory devices.

S11. Pohm, A. V., Huang, J. S. T., Daughton, J. M., Krahn, D. R. \& Mehra, V. The design of a one megabit nonvolatile M-R memory chip using 1.5x5 $\mu$m cells. \emph{IEEE Trans. Magn.}
\textbf{24}, 3117 (1988).

S12. Fert, A., Cros, V. \& Sampaio, J. Skyrmions on the track. \emph{Nature Nanotech.} \textbf{8}, 152 (2013).

S13. Son, Y. H., O, S., Ro, Y., Lee, J. W. \& Ahn, J. H. Reducing memory access latency with asymmetric DRAM bank organizations. \emph{Proceedings of the 40th Annual International Symposium on Computer Architecture}. June 23-27, Tel-Aviv, Isreal (2013).

S14. Parkin, S. S. P., Hayashi, M. \& Thomas, L. Magnetic domain-wall racetrack memory. \emph{Science} \textbf{320}, 190 (2008).

\end{document}